\documentclass[lettersize,journal]{IEEEtran}
\usepackage{amsmath,amsfonts}
\usepackage{algorithmic}
\usepackage{array}
\usepackage[caption=false,font=normalsize,labelfont=sf,textfont=sf]{subfig}
\usepackage{textcomp}
\usepackage{stfloats}
\usepackage{url}
\usepackage{verbatim}
\usepackage{graphicx}
\usepackage{optidef}
\usepackage[caption=false]{subfig}
\usepackage{supertabular, booktabs}

\usepackage[style=numeric,sorting=none]{biblatex}
\usepackage{balance}

\begin{document}
\title{Improving Data Cleaning Using Discrete Optimization}

\author{\IEEEauthorblockN{Kenneth Smith\IEEEauthorrefmark{1} and
Sharlee Climer \IEEEauthorrefmark{1}\IEEEauthorrefmark{2}} \\
\IEEEauthorblockA{\IEEEauthorrefmark{1}Department of Computer Science,
University of Missouri - St. Louis, St. Louis, MO 63121 USA} \\
\IEEEauthorblockA{\IEEEauthorrefmark{2}Corresponding Author: climer@umsl.edu}}


\maketitle

\begin{abstract}
One of the most important processing steps in any analysis pipeline is handling missing data. Traditional approaches simply delete any sample or feature with missing elements. Recent imputation methods replace missing data based on assumed relationships between observed data and the missing elements. However, there is a largely under-explored alternative amid these extremes. Partial deletion approaches remove excessive amounts of missing data, as defined by the user. They can be used in place of traditional deletion or as a precursor to imputation. In this manuscript, we expand upon the Mr. Clean suite of algorithms, focusing on the scenario where all missing data is removed. We show that the RowCol Integer Program can be recast as a Linear Program, thereby reducing runtime. Additionally, the Element Integer Program can be reformulated to reduce the number of variables and allow for high levels of parallelization. Using real-world data sets from genetic, gene expression, and single cell RNA-seq experiments we demonstrate that our algorithms outperform existing deletion techniques over several missingness values, balancing runtime and data retention. Our combined greedy algorithm retains the maximum number of valid elements in 126 of 150 scenarios and stays within 1\% of maximum in 23 of the remaining experiments. The reformulated Element IP complements the greedy algorithm when removing all missing data, boasting a reduced runtime and increase in valid elements in larger data sets, over its generic counterpart. These two programs greatly increase the amount of valid data retained over traditional deletion techniques and further improve on existing partial deletion algorithms. 
\end{abstract}


\section{Introduction} \label{sec:intro}
\IEEEPARstart{T}{he} first step in most data processing pipelines is data cleaning. This preparatory procedure involves removing duplicate data, along with the handling of erroneous and missing data. The widespread occurrence of missing data across different domains (gene expression analysis \cite{FarswanA2020, DubeyA2021, Souto2015}, psychology \cite{PeetersM2015}, epidemiology \cite{MadleyDowd2019}, genetics \cite{Petrazzini2021}), combined with the inability of many analysis algorithms to function correctly with missing data, makes data cleaning a critical area of research.
\par
In his seminal work, Rubin outlined three mechanisms behind missing data \cite{RubinD1975}. These mechanisms are differentiated by whether or not the probability of an element being missing is independent, dependent on observed data, or dependent on the missing value itself. If the probability of a given element of data being missing is independent of both the value of the data and the other data elements, then it is considered Missing Completely at Random (MCAR). When the probability of an element being missing is independent of the missing value, but dependent at least in part on other observed values, it is called Missing at Random (MAR). This is occasionally observed in psychology studies when the probability of an individual answering a question is related to their gender. The third mechanism, called Missing Not at Random (MNAR), occurs when the probability of an element being missing is based on the value of the missing data. Missing data can occur in gene expression studies, when values below the minimum detection threshold are considered missing. In this case, the actual value of the data impacts its probability to be considered missing. Despite work in understanding the mechanisms behind missing data, it is not always possible to determine which mechanism or combination of mechanisms is causing missing data in a given experiment. However, it is important to understand the biases introduced to a data set by various cleaning algorithms, based on the missingness mechanisms.
\par
Missing data is generally handled either by deletion or imputation. Deletion techniques remove samples or features from a data set in order to delete missing elements. The simplest version of deletion, and the default in many software packages at one time \cite{Briggs2003}, is \textit{list-wise} deletion, where a sample is deleted if it contains any missing data. Testing power is reduced using this technique, as the number of samples is decreased. Additionally, if the data is not MCAR, \textit{list-wise} deletion results in bias \cite{NewmanD2014}. Similarly, \textit{feature-wise} deletion removes any feature with missing data. This technique keeps the original testing power of the experiment, but risks deleting strongly predictive features and introducing biases of its own \cite{vanGinkel2020}. Another common deletion techniques is called \textit{pair-wise} deletion, where samples are deleted if they contain any missing data in a subset of the features. Consider the scenario where a regression model is to be created using three features. When cleaning with \textit{pair-wise} deletion, samples are removed only if missing data occurs in those three features, regardless of missing data in the remaining features. While \textit{pair-wise} deletion often retains more samples than \textit{list-wise}, it can result in a covariance matrix that is not positive definite, causing issues for some techniques \cite{NewmanD2014}.
\par
So far, we have only discussed deletion techniques that remove all missing data. However, there is a family of \textit{partial deletion} algorithms which remove excessive missing data, as defined by the user. \textit{DataRetainer} \cite{Climer2014, ClimerS2021} is an interactive algorithm, where the user specifies the maximum percentage of missing data allowed in each retained row and column, separately. All rows and columns with too much missing data can be removed all at once, or in an iterative fashion. \textit{Auto-miss} \cite{AutoMiss} is a greedy algorithm where the user provides an allowed amount of missing data for the entire data set. Rows and columns are removed until the cleaned matrix contains the desired amount of missing data. The \textit{Mr. Clean} suite \cite{MrClean} contains three algorithms which delete rows and columns based on a user specified maximum percentage of missing data. Unlike \textit{Auto-miss}, this percentage applies to each row and column in the cleaned matrix. The \textit{Element} Integer Program (IP) within \textit{Mr. Clean} retained the optimal number of elements in small data sets, but was unable to converge for larger data sets. The \textit{greedy} algorithm was able to easily process all of the data sets and retain $>99\%$ of the optimal number of elements in each experiment. These programs showed a substantial increase in data retention over existing \textit{partial deletion} algorithms without the need for user interaction.
\par
While deletion techniques remove data from an experiment, imputation replaces missing data based on observed values and assumed relationships. Mean imputation is the simplest technique and involves replacing missing data with the mean, on a feature-by-feature basis. This can be performed on the raw data or after normalizing the data set. In either case, the variance and covariance are biased downward \cite{vanGinkel2020}. Local methods impute data based on similar objects according to a similarity measures, such as L2 norm or Pearson's correlation \cite{DubeyA2021, Kim2005, Troyanskaya2001}. Souto \cite{Souto2015} found that more advanced local similarity algorithms had negligible improvements in gene expression clustering, compared to simpler techniques. Numerous other techniques have been proposed including Bayesian principal component analysis (BPCA) \cite{Oba2003}, expectation maximization  (EMimute) \cite{Bo2004}, and several variants of  regression imputation \cite{Souto2015, Lin2020, Musil2002}.
\par
Regression imputation tries to capitalize on any relationships between missing values and observed features, by training models to predict missing values using the observed data as predictors. Some error may be introduced if the model does not accurately capture the dependence between features. Additionally, the variance for imputed features is biased downward, while the covariance is biased upwards \cite{vanGinkel2020}. Random error terms are added to regression imputation to counter act the variance and covariance biases. This advanced regression imputation, called stochastic regression, suffers from a bias in the p-values and confidence intervals \cite{vanGinkel2020}. The most robust variant of imputation is Multiple Imputation (MI) \cite{vanGinkel2020}. In MI, multiple stochastic regression models, typically 5 - 20 \cite{Austin2021}, are generated on the target data set. Each of these imputed data sets are independently analyzed and the results are combined by various pooling techniques. When an accurate imputation model is selected, the MI estimates are unbiased for MCAR and MAR data. The major drawback of MI is the increased computational requirements, which may not always outweigh the benefits \cite{WoodsAD2021}. When the data is MNAR, MI can still suffer biases, but these can be reduced by using the missing indicator method \cite{ChoiJ2019}. Imputation methods, in general, can reinforce patterns in observed data and inflate correlation structures \cite{ClimerS2021}. Thus, despite the wide range of available deletion and imputation techniques, no consensus has been reached on the best way to handle missing data.
\par
We continue this manuscript with an overview of the \textit{Mr. Clean} algorithms in Section \ref{sec:related}, focusing on the two IPs introduced in \cite{MrClean}. In Section \ref{sec:methods}, we add our contribution to this field, by considering the impact to the \textit{Mr. Clean} IPs when no missing data is allowed. By restricting ourselves to this scenario, we will show that both \textit{Mr. Clean} IPs can be recast to dramatically improve processing time and memory requirements. Additionally, we will develop a new greedy algorithm designed specifically for the no missing data allowed scenario. In Section \ref{sec:results}, we compare the \textit{Mr. Clean} algorithms to their new counterparts and then to several existing partial deletion programs using 50 real-world data sets.

\section{Related Work} \label{sec:related}
\textit{Mr. Clean} details three \textit{partial deletion} algorithms aimed at maximizing the number of valid elements in a cleaned data matrix, while requiring that the percentage of missing data in each retained row and column is no more than $\gamma \in [0,1)$. This problem can be expressed mathematically by considering the $m \times n$ data matrix $\mathbf{D}$, where $d_{ij}$ is the element in the $i$-th row and $j$-th column of $\mathbf{D}$. An element is retained in the data matrix if the corresponding row and column are kept. The decision variables $r_i$ are used to indicate which rows are kept, while the variables $c_j$ are used for columns. If $r_i=0$, then all elements from row $i$ are removed in the cleaned matrix. Likewise, $c_j=0$ indicates all elements in column $j$ are removed.

\subsection{Greedy} \label{sec:related_greedy}
The first algorithm presented in \textit{Mr. Clean} is a \textit{greedy} algorithm. A \textit{greedy} algorithm is an approximation algorithm that makes the locally \textit{best} decision at each iteration or decision point. These are typically fast running algorithms, but may result in poor solutions. The greedy algorithm in \textit{Mr. Clean} begins by initializing all $r$ and $c$ decision variables to 1. It then iteratively removes rows and/or columns until the percentage of missing data in all remaining rows and columns is no more than $\gamma$. At each iteration, the row or column with the largest percentage of missing data is identified. Next, the algorithm calculates $k$, the number of invalid elements, in the selected row or column, that need to be removed in order for the percentage of missing data to be $\leq \gamma$. Assuming a row contains the largest percentage of missing data, the number of valid elements in that row is compared to the minimum number of valid elements in every combination of $k$ columns, where the columns must contain missing data in the desired row. If the row contains more valid elements, the columns are removed. If, on the other hand, the columns contain more valid elements, the row is removed. A similar approach is taken if a column contains the largest percent of missing elements. The pseudo-code for this algorithm can be found in \cite{MrClean}.

\subsection{RowCol IP} \label{sec:related_rowcol}
The second algorithm in the \textit{Mr. Clean} suite is the \textit{RowCol} IP. IPs are mathematical optimization programs where all of the decision variables are restricted to integers. Additionally, the IPs in \textit{Mr. Clean} contain only linear objective functions and linear constraints. From \cite{MrClean}, the \textit{RowCol} IP is formulated as:
\begin{maxi!}|l|[3]
{r,c}{\sum_{i=1}^{m} \alpha_{i} r_{i} + \sum_{j=1}^{n} \beta_{j} c_{j}}
{}{} \label{rowcolobj}
\addConstraint{r_i + \sum_{j=1}^{n} \left ( \frac{1 - b_{ij} - \gamma}{n} \right ) c_j \leq 1}{\quad}{ \forall \; i} \label{rowcolip:row}
\addConstraint{c_{j} + \sum_{i=1}^{m} \left ( \frac{1 - b_{ij} - \gamma }{m} \right ) r_{i} \leq 1}{\quad}{\forall \; j} \label{rowcolip:col}
\addConstraint{r_i,c_j \in \{ 0,1 \}}{\quad}{\forall \; i,j} \label{rowcolip:intergral}
\end{maxi!}
where $m$  and $n$ are the number of rows  and columns in the original data matrix, respectively;  $\alpha_i$ and $\beta_j$ are the number of valid elements in the $i$-th row and $j$-th column of the original data matrix, respectively; $\gamma$ is the maximum percentage of missing data allowed in each row and column of the cleaned matrix expressed as a decimal, $b_{ij}$ indicates if $d_{ij}$ is valid in the original data matrix, and $r_i$ and $c_j$ are decision variables indicating if row $i$ and column $j$, respectively, are kept in the solution. It is important to note that the objective function in Eq. \eqref{rowcolobj} is based on the number of valid elements in each row and column in the original data matrix. However, as rows and columns are removed, the number of valid elements in the remaining rows and columns will change. The \textit{RowColFail} example in \cite{MrClean} demonstrates that optimizing this IP does not guarantee maximizing the number of valid elements in the cleaned matrix. But in general, the difference in the number of valid elements in the cleaned data matrix between the two methods is small, while \textit{RowCol} IP is significantly faster than the \textit{Element} IP.

\subsection{Element IP} \label{sec:related_element}
In order to account for the changing number of valid elements as rows and columns are removed, the \textit{Element} IP was developed, as detailed in \cite{MrClean}. Accounting for the dynamic number of valid elements was accomplished by introducing additional decision variables $x_{ij}$, which indicate if element $d_{ij}$ is kept. An element is kept only when the corresponding row and column are both kept. Variables in common with \textit{RowCol} IP have the same meaning. If this IP is solved to optimality, the solution is guaranteed to retain the maximum number of valid elements, given the $\mathbf{D}$ and $\gamma$.
\begin{maxi!}|l|[3]
{x}{\sum_{i=1}^{m} \sum_{j=1}^{n} b_{ij} x_{ij}}
{}{} \label{element:obj}
\addConstraint{x_{ij} \leq \frac{1}{2} \left( r_{i} + c_{j} \right )}{\quad}{\forall \; i,j} \label{element:x}
\addConstraint{r_i + \sum_{j=1}^{n} \left ( \frac{1 - b_{ij} - \gamma}{n} \right ) c_j \leq 1}{\quad}{ \forall \; i} \label{element:row}
\addConstraint{c_{j} + \sum_{i=1}^{m} \left ( \frac{1 - b_{ij} - \gamma }{m} \right ) r_{i} \leq 1}{\quad}{\forall \; j} \label{element:col}
\addConstraint{x_{ij},r_i,c_j \in \{ 0,1 \}}{\quad}{\forall \; i,j} \label{element:integer}
\end{maxi!}

\section{Methods} \label{sec:methods}
In this section we reformulate the two IPs from \textit{Mr. Clean} for the special case where $\gamma=0$. Then we introduce a new \textit{greedy} algorithm designed for this special case.

\subsection{RowCol LP} \label{methods:rowcol}
To begin, we simplify Eqs. \eqref{rowcolip:row} and \eqref{rowcolip:col} by substituting in $\gamma=0$.
\begin{align*}
    r_i + \sum_{j=1}^{n} \left ( \frac{1 - b_{ij}}{n} \right ) c_j \leq 1 \quad \forall i \\
    c_{j} + \sum_{i=1}^{m} \left ( \frac{1 - b_{ij}}{m} \right ) r_{i} \leq 1 \quad \forall j
\end{align*}
To better understand the relationship between $r_i$ and $c_j$, consider the example in which  $b_{11}=0$ and $r_1=1$. The constraints for row 1 and column 1 can be rewritten as:
\begin{align*}
    1 + \frac{1}{n} c_1 + \sum_{j=2}^{n} \left ( \frac{1 - b_{ij}}{n} \right ) c_j \leq 1\\
    c_1 + \frac{1}{m}  + \sum_{i=2}^{m} \left ( \frac{1 - b_{ij}}{m} \right ) r_{i} \leq 1
\end{align*}
Since the value of both summations is non-negative and $c_1 \in \{0,1\}$, according to both constraints $c_1=0$. Similarly, if $b_{11}=0$ and $c_1=1$, then $r_1=0$. This can be generalized to any invalid element in the matrix. In other words, if an element contains invalid or missing data, then the corresponding row and column cannot both be kept. This requirement can be implemented by replacing the \textit{RowCol} constraints with the following constraint for each invalid element in the data matrix.
\begin{equation}
    r_{i} + c_{j} \leq 1 \quad \forall \; (i,j) | b_{ij}=0
    \label{noMissRowCol}
\end{equation}

These new \textit{RowCol} constraints can be represented as the system of linear inequalities $\mathbf{A}x \leq \mathbf{b}$, where $x^{T} = [r_1,r_2,\dots,r_m,c_1,c_2,\dots,c_n]$, $\mathbf{b_i}=1 \; \forall i$, and each row of $\mathbf{A}$ contains 1's corresponding to the row and column of a single missing element in $\mathbf{D}$. Interestingly, $\mathbf{A}^{T}$ is the undirected incidence matrix of a bipartite graph. This matrix, along with its transpose, are totally unimodular (TU). Since $\mathbf{A}$ is TU and $\mathbf{b}$ is integral, the optimal solution of the Linear Program (LP) constrained by $\mathbf{A}x \leq \mathbf{b}$ is integral \cite{Saaty1970}. Thus, we can remove the integer constraints and still obtain an integer solution. Therefore, the original \textit{RowCol} IP can be recast as the following LP:
\begin{maxi!}|l|[3]
{r,c}{\sum_{i=1}^{m} \alpha_{i} r_{i} + \sum_{j=1}^{n} \beta_{j} c_{j}}
{}{}
\addConstraint{r_i + c_j \leq 1}{\quad}{\forall \; (i,j)|b_{ij}=0}
\end{maxi!}

\subsection{Element IP} \label{methods:element}
We can simplify the \textit{Element} IP by replacing the Eqs. \eqref{element:row} and \eqref{element:col} constraints with Eq. \eqref{noMissRowCol} as above, when considering the $\gamma=0$ scenario. However, the resulting matrix is not TU, due to the constraints involving $x_{ij}$. We found that solving this new IP intractable for large data sets.
\par
In order to execute the \textit{Element} IP on large data sets, we needed to reduce the computation complexity, either by reducing the number of variables, modifying the constraints, or distributing the problem over several processors. The \textit{Element} IP can easily be split into sub-problems, based on the number of rows in the cleaned data matrix. Let us partition the \textit{Element} IP solution space, based on the sum of the $r_i$ decision variables. An IP that searches one of these partitions is created by adding the following constraint:
\begin{equation}
    \sum_{i=1}^{m} r_{i} = R
\end{equation}
If a sub-problem is searched for each valid value of $R$, then the entire \textit{Element} IP solution space will also be searched. However, adding this constraint to the \textit{Element} IP and solving still results in a large number of unnecessary constraints and variables.
\par
When $\gamma=0$, any matrix satisfying the \textit{Element} IP contains no invalid elements. Therefore, the number of valid elements in the cleaned matrix is equal to the number of rows in the cleaned matrix multiplied by the number of columns. For a given number of rows, the number of elements increases as the number of columns increases, with the maximum number of elements corresponding to the maximum number of columns. For each sub-problem constrained with $R$, the \textit{Element} IP objective function can be replaced by the sum of the column decision variables. Since the $x_{ij}$ variables no longer appear in the objective function, nor do they restrict other variables, the corresponding constraints can be dropped from the problem. This reduces the number of variables from $(m + n + m*n)$ to $(m + n)$. Finally, we replace the original row and column constraints with a constraint for each missing element, resulting in a new \textit{MaxCol} IP.
\begin{maxi!}|l|[3]
{c,r}{\sum_{j=1}^{n} c_{j}} 
{}{}\label{maxCol:obj}
\addConstraint{r_{i} + c_{j} \leq 1}{\quad}{\forall \; (i,j) | b_{ij}=0} \label{maxcol:missing}
\addConstraint{\sum_{i=1}^{m} r_{i} = R}
\addConstraint{r_i,c_j \in \{ 0,1 \}}{\quad}{\forall \; i,j}
\end{maxi!}

\subsubsection{Reducing the Number of Constraints}
This implementation of \textit{MaxCol} works well for small and medium size data sets, but still proves difficult for large data sets, due in part to the size of the constraint matrix. We can reduce the number of constraints to $\leq (m + 1)$, by combining all of the missing element constraints for each row into a single constraint. Let $\lambda_i = n - \sum_{j=1}^{n}{b_{ij}}$, be the number of missing elements in row $i$. If any of the columns corresponding to a missing element in row $i$ is kept, then $r_i$ must equal 0. This can be accomplished via the following constraint.
\begin{equation*}
    r_i + \frac{1}{\lambda_i} \sum_{j=1}^{n}{c_j(1-b_{ij})} \leq 1
\end{equation*}
The maximum value of the sum is $\lambda_i$, which results in $r_i \leq 0$. Any positive summation less than $\lambda_i$, will result in $r_i < 1$. This combined with the integer constraint on $r_i$ will force it to 0. Finally, when all columns corresponding to missing data are removed, the inequality become $r_i \leq 1$, allowing the row to be kept. If a row contains no missing data, then $\lambda_{i} = 0$, and dividing by $\lambda_i$ is undefined. We can avoid this undefined behavior by only applying the constraint when $\lambda_i > 0$.

\subsubsection{Reducing the Number of Integer Variables}
As shown above, any positive value of $c_j$ corresponding to a column with missing data in row $i$ will force $r_i$ to 0. But if $c_j > 0$, the objective function will be maximized when $c_j = 1$. Thus $c_j \in \{0,1\}$ without the need to add integer constraints for these variables. Thus, we arrive at our final \textit{MaxCol} MIP formulation.
\begin{maxi!}|l|[3]
{c,r}{\sum_{j=1}^{n} c_{j}}
{}{}
\addConstraint{\sum_{i=1}^{m} r_{i} = R}
\addConstraint{r_i + \frac{1}{\lambda_i} \sum_{j=1}^{n}{c_j(1-b_{ij})} \leq 1}{\quad}{\forall \; i|\lambda_i > 0} \label{element:largedata}
\addConstraint{r_i \in \{ 0,1 \}}{\quad}{\forall \; i}
\addConstraint{c_j \in  [0,1]}{\quad}{\forall \; j}
\end{maxi!}

\subsubsection{Removing Variables}
One way to further reduce the number of variables in each sub-problem, and possibly reduce the gap between the LP relaxation and the optimal integer solution, is to remove variables that cannot be in the optimal solution. A column can only be kept in a \textit{MaxCol} solution if it contains at least $R$ valid elements. Therefore, we can remove all columns with less than $R$ valid elements from each sub-problem.

Similarly, we can remove rows that contain less valid elements than required. In order to determine the minimum number of elements required per row, or minimum number of columns in the solution, we turn to an incumbent solution. This solution can be provided from a different algorithm or from an earlier sub-problem. Let $obj^*$ be the total number of valid elements in the current incumbent solution. For any sub-problem, the minimum number of columns required for a better solution is calculated as:
\begin{equation} \label{eq:minC_calc}
    minC = \left \lfloor \frac{obj^*}{R} \right \rfloor + 1
\end{equation}
When $obj^*$ is a multiple of $R$, $\left \lfloor \frac{obj^*}{R} \right \rfloor = \frac{obj^*}{R}$. If we didn't add 1 to the minimum number of columns, a solution with the same objective value could be found. If $obj^*$ is not a multiple of $R$, the addition of 1 prevents the possibility of a solution with a smaller objective value from being found. Any row containing fewer valid elements than $minC$ is removed from the sub-problem.

We also calculate the number of mutually valid elements in each pair of rows. For each pair of rows $i$ and $i'$, where $i \neq i'$, we count the number of columns where elements $b_{ij}=1$ and $b_{i'j}=1$. If the number of columns is less than $minC$, then rows $i$ and $i'$ cannot both be in the optimal solution. If enough row-pairs containing row $i$ are removed, then row $i$ cannot be an any optimal solution. In particular, let $\sigma_i = \{i' | rowPairCount(i,i') \geq minC, i' \neq i\}$. If $|\sigma_i| < R-1$, then inclusion of row $i$ in a solution will leave less than $R$ valid rows. Since this cannot be a possible solution to the sub-problem, we remove row $i$. Additionally, we add $minC$ as a lower bound to the MIP.

\subsubsection{Skipping Sub-problems}
In certain cases, the MIP portion of a sub-problem can be skipped based on the $R$ and $minC$ values. First, if $minC > n$  for any sub-problem, then the MIP can be skipped. Also, if the number of remaining valid rows $ < R$ or the number of remaining valid columns is $< minC$, the sub-problem can also be skipped.

\subsection{NoMiss Greedy}
The greedy algorithm in this manuscript utilizes the property that if $d_{ij}$ is missing, then row $i$ and column $j$ cannot both be kept. The greedy solution is initialized so that all columns are kept and all rows are removed. Next, the \textit{best} row is selected and added to the solution. If the selected row contains any missing elements, the corresponding column(s) are removed from the solution. The objective value is calculated by multiplying the number of kept rows by the number of kept columns. This process is repeated until all rows have been selected. The kept rows and columns corresponding to the best objective value are returned as the greedy solution.
\par

At each iteration, the \textit{best} row is selected in a two-step process. First, the number of valid elements in each row is calculated. Only elements in retained columns are considered. From rows not currently in the solution, the maximum number of valid elements is found. If only one row contains the maximum number of valid elements, it is selected as the best row. However, if multiple rows contain the maximum amount of valid elements, then the second step is used to select the \textit{best} row. For each row with the maximum number of valid elements, the number of missing elements that would be removed by its inclusion is counted. Only elements contained by rows with a similar number of valid elements are considered. In the experiments presented below, the number of elements in similar rows could only differ by 3. The row that removes the most missing elements is selected. If a tie occurs after the second check, the first row found is selected.

\section{Results} \label{sec:results}
We compared the performance of our new algorithms to the original \textit{Mr. Clean}, \textit{Auto-miss}, \textit{list-wise}, \textit{feature-wise}, \textit{naive}, and \textit{DataRetainer} algorithms, using 50 real-world data sets. Discrete data sets were obtained from chromosome 1, 17, 19, and Y from all 11 populations of the HapMap3 project \cite{HapMapIII}. A variety of continuous data sets were gathered, including the kamyr-digest set \cite{DunnK2011, DayalB1994}, gene expression experiments \textit{GSE54456} \cite{Li2014}, \textit{GSE215307} \cite{SzaboP2023} and \textit{AD\_CSF} \cite{YangC2021}, and single cell RNA-seq values from GSE146026 \cite{Izar2020} and GSE146264 \cite{Liu2021}. Each of the 50 data sets were oriented so that the number of rows was smaller than the number of columns, as this improved the effectiveness of the \textit{greedy} algorithms and reduced runtime for the \textit{MaxCol} MIP, without impacting the other algorithms. Details of the original data sets are shown in Table \ref{tab:data_sets}. 

\begin{table}[ht]
\begin{center}
\caption{Summary of Data Sets}
\label{tab:data_sets}
\begin{tabular}{ccccc}
\hline
ID & Name & \# Rows & \# Cols & \% Missing \\
\hline
1 & kamyr-digest & 301 & 22 & 5.3 \\
2 & ChrY-MEX & 605 & 86 & 0.8 \\
3 & ChrY-ASW & 616 & 87 & 0.5 \\
4 & ChrY-TSI  & 556 & 102 & 2.4 \\
5 & ChrY-GIH & 606 & 101 & 0.7 \\
6 & ChrY-CHD & 576 & 109 & 0.8 \\
7 & ChrY-LWK & 620 & 110 & 1.1 \\
8 & ChrY-JPT & 926 & 116 & 32.1 \\
9 & ChrY-MKK & 585 & 184 & 1.0 \\
10 & ChrY-CHB & 930 & 139 & 32.4 \\
11 & ChrY-CEU & 943 & 174 & 31.2 \\
12 & ChrY-YRI & 922 & 209 & 30.2 \\
13 & AD\_CSF & 489 & 705 & 2.2 \\
14 & Chr19-MEX & 26,888 & 86 & 0.4 \\
15 & Chr19-ASW & 27,866 & 87 & 0.3 \\
16 & Chr19-GIH & 25,389 & 101 & 0.3 \\
17 & Chr19-CHD & 23,876 & 109 & 0.4 \\
18 & Chr19-TSI & 25,665 & 102 & 0.3 \\
19 & Chr19-LWK & 27,869 & 110 & 0.4 \\
20 & Chr17-MEX & 38,280 & 86 & 0.4 \\
21 & Chr17-ASW & 40,768 & 87 & 0.3 \\
22 & GSE54456 & 174 & 21,099 & 12.5 \\
23 & Chr17-CHD & 34,004 & 109 & 0.4 \\
24 & Chr17-GIH & 36,827 & 101 & 0.3 \\
25 & Chr17-TSI & 37,391 & 102 & 0.3 \\
26 & Chr17-LWK & 40,529 & 110 & 0.4 \\
27 & Chr19-MKK & 27,525 & 184 & 0.4 \\
28 & Chr19-JPT & 59,520 & 116 & 38.4 \\
29 & Chr17-MKK & 40,422 & 184 & 0.4 \\
30 & Chr19-CHB & 59,571 & 139 & 41.0 \\
31 & Chr1-MEX & 117,736 & 86 & 0.5 \\
32 & Chr19-CEU & 59,707 & 174 & 30.7 \\
33 & Chr17-JPT & 92,177 & 116 & 40.4 \\
34 & Chr1-ASW & 125,129 & 87 & 0.3 \\
35 & Chr1-GIH & 112,989 & 101 & 0.3 \\
36 & Chr1-CHD & 104,948 & 109 & 0.4 \\
37 & Chr1-TSI & 113,882 & 102 & 0.3 \\
38 & Chr19-YRI & 58,839 & 209 & 32.8 \\
39 & Chr17-CHB & 92,266 & 139 & 43.1 \\
40 & Chr1-LWK & 123,617 & 110 & 0.4 \\
41 & Chr17-CEU & 92,433 & 174 & 31.9 \\
42 & Chr17-YRI & 90,944 & 209 & 34.3 \\
43 & GSE215307 & 375 & 57,905 & 25.5 \\
44 & Chr1-MKK & 124,198 & 184 & 0.4 \\
45 & Chr1-JPT & 317,446 & 116 & 42.7 \\
46 & Chr1-CHB & 317,642 & 139 & 45.8 \\
47 & Chr1-CEU & 314,222 & 174 & 33.6 \\
48 & Chr1-YRI & 313,459 & 209 & 36.8 \\
49 & GSE14602 & 11,548 & 9,609 & 77.0 \\
50 & GSE146264 & 58,129 & 7,303 & 96.1 \\
    \hline
    \end{tabular}
\end{center}
\end{table}

\subsection{Greedy Algorithms}
\begin{figure}[htbp]
    \subfloat[\label{fig:greedy_elements}]{
        \includegraphics[width=0.8\linewidth]{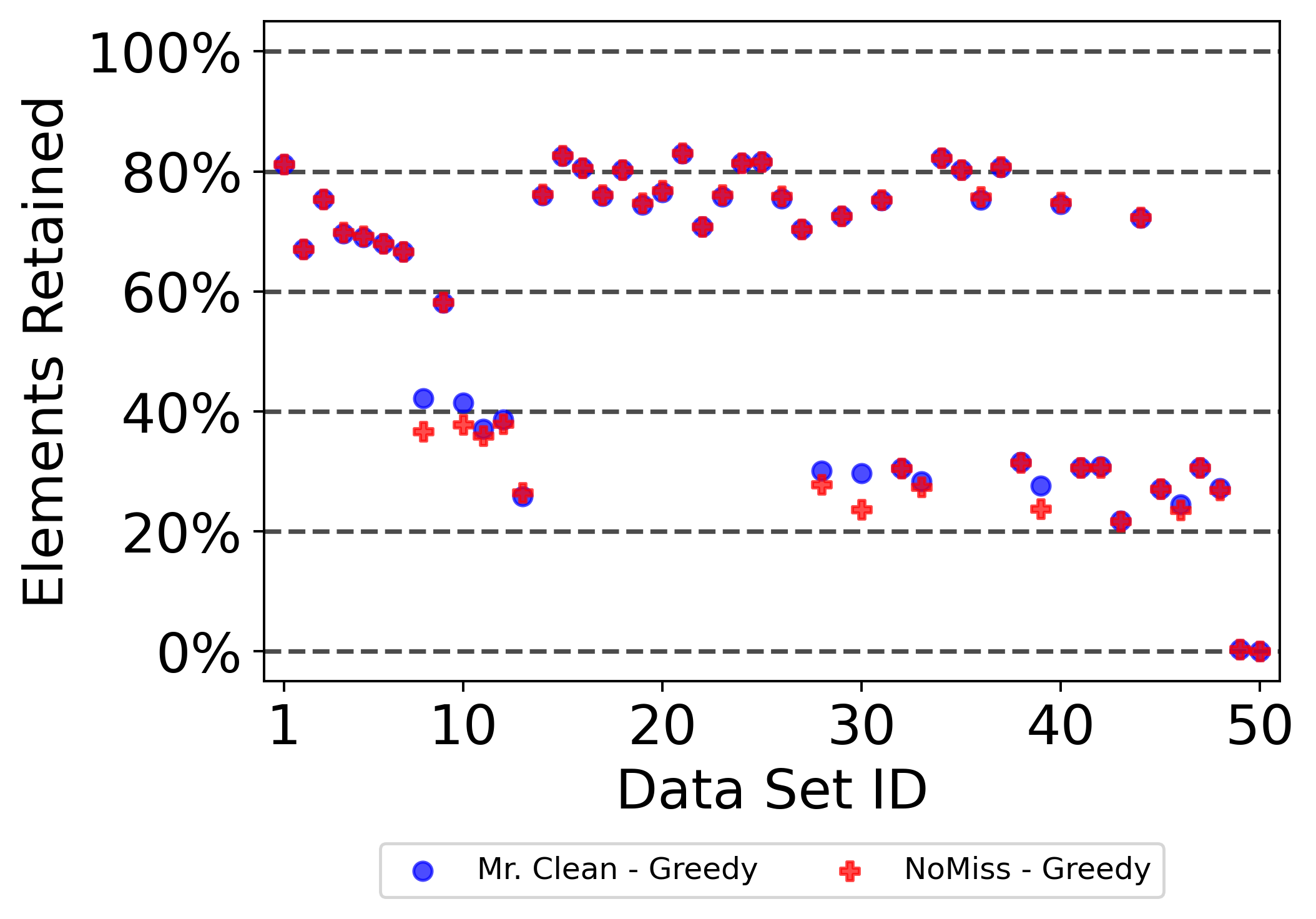}
    }
    \hfill
    \subfloat[\label{fig:greedy_runtime}]{
        \includegraphics[width=0.8\linewidth]{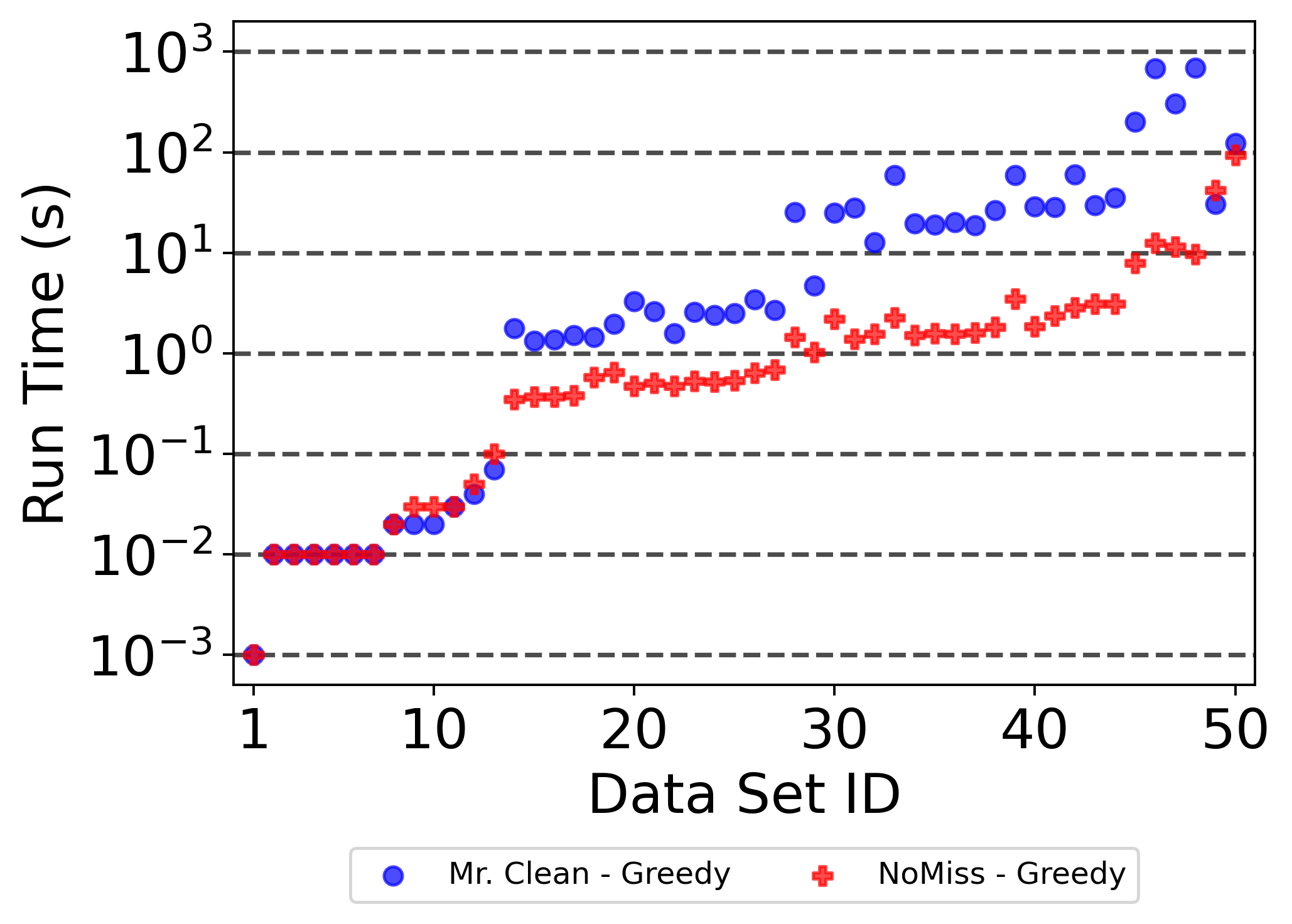}
    }
    \hfill
    \subfloat[\label{fig:greedy_rows}]{
        \includegraphics[width=0.8\linewidth]{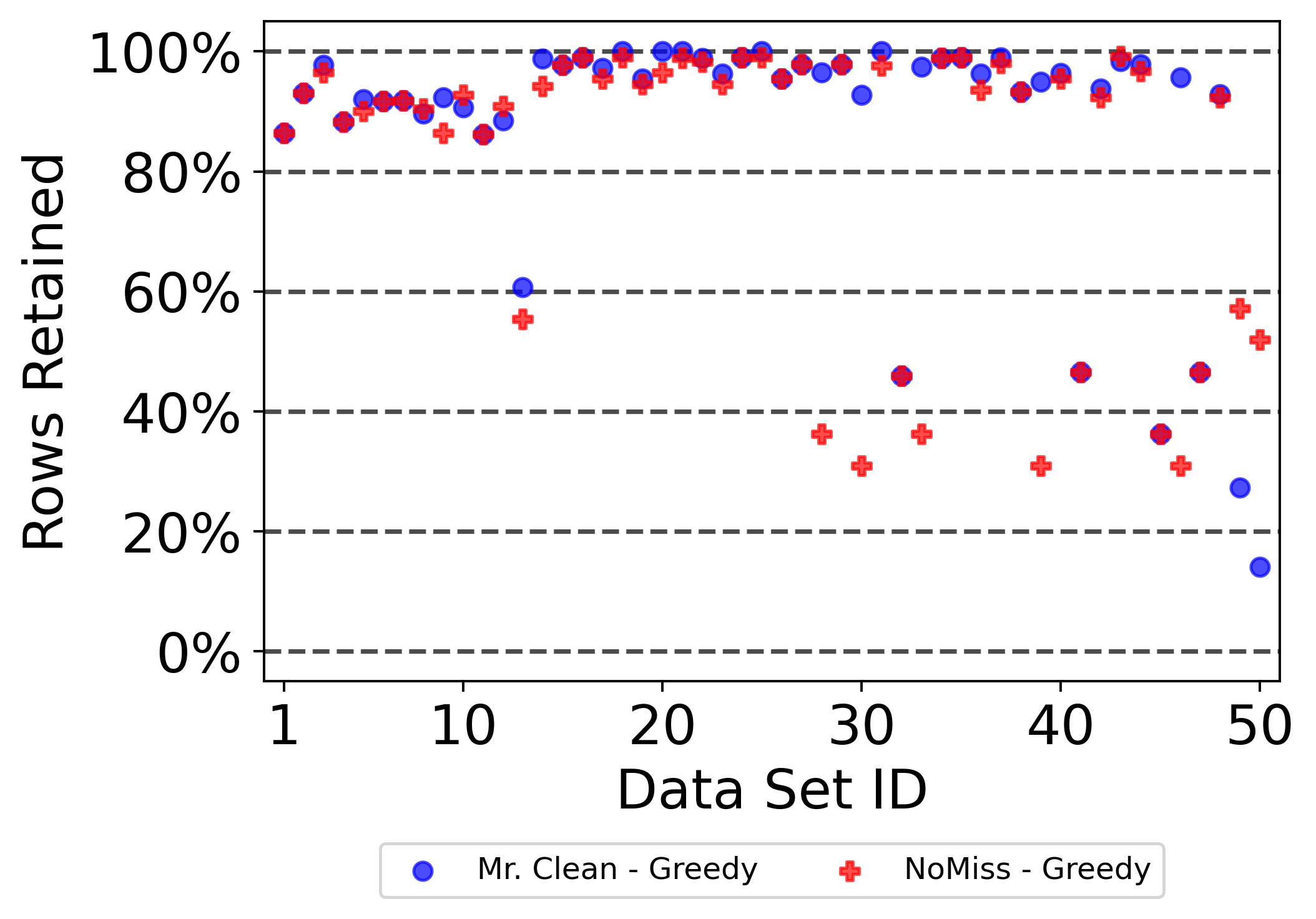}
    }
    \hfill
    \subfloat[\label{fig:greedy_cols}]{
        \includegraphics[width=0.8\linewidth]{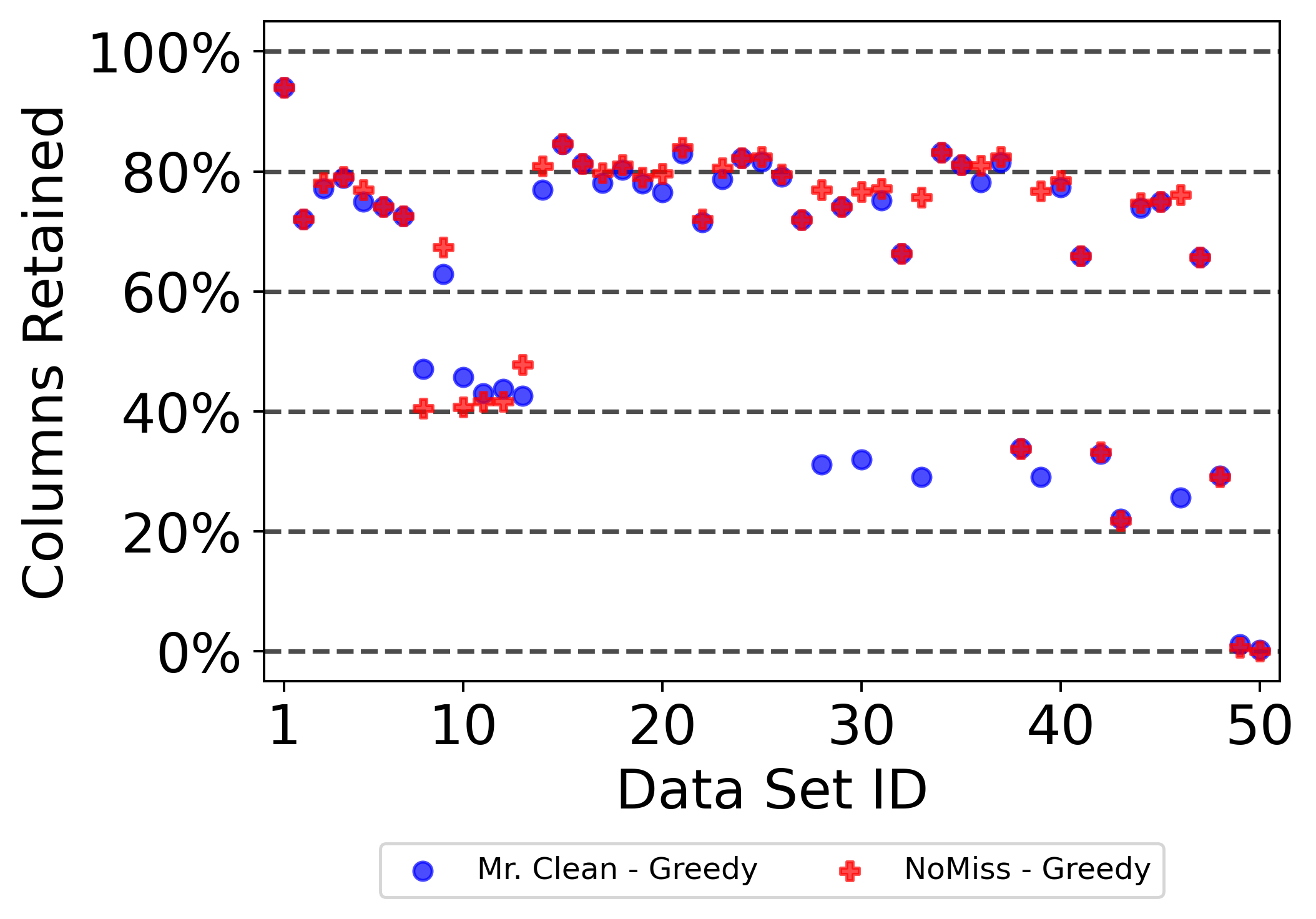}
    }
    \caption{Comparison of \textit{Mr. Clean} and \textit{NoMiss} Greedy Algorithms. \textbf{(a)} Percentage of Original Valid Elements Retained \textbf{(b)} Run Time. \textbf{(c)} Percentage of Original Rows Retained. \textbf{(d)} Percent of Original Columns Retained.} 
    \label{fig:greedy}
\end{figure}

We begin our analysis by comparing the \textit{Mr. Clean - Greedy} algorithm with our new \textit{NoMiss - Greedy} algorithm (Figure \ref{fig:greedy}). Figure \ref{fig:greedy_elements} shows the percent of valid elements retained by each algorithm. In 24 of the 50 trials, \textit{NoMiss - Greedy} retained slightly more elements than \textit{Mr. Clean - Greedy}. However, with the exception of data set 13, these improvements were less than 0.5\%. In 14 of the 50 trials, \textit{NoMiss - Greedy} retained less valid elements than \textit{Mr. Clean - Greedy}, varying by almost 6\% for data set 30.

The \textit{NoMiss - Greedy} algorithm was substantially faster than \textit{Mr. Clean - Greedy} in 35 of the 50 runs, as seen in Figure \ref{fig:greedy_runtime}. Both algorithms had the same run time (to within 0.001 seconds) for the first eight trials. However, \textit{Mr. Clean - Greedy} quickly took an order of magnitude, or more, longer to run than \textit{NoMiss - Greedy.} Luckily, the absolute run time of both algorithms was very quick, with the longest combined run time of $\sim11.5$ minutes, occurring at data sets 46 and 48. Since the \textit{No Miss} implementation had a quick run time, we decided to run both greedy algorithms when $\gamma=0$ and take the best solution. This approach is used in the rest of this manuscript.

In Figures \ref{fig:greedy_rows} and \ref{fig:greedy_cols}, we compared the percent of rows and columns retained by the two algorithms. The two algorithms retained the same number of rows in 19 data sets, but the same number of columns in only 12. These 12 instances were the only ones where the two algorithms returned a matrix with the same number of valid elements. In general, \textit{NoMiss - Greedy} removed more rows and kept more columns. This is especially evident in a few instances near trials 30 and 40, where \textit{NoMiss - Greedy} kept $\sim30\%$ of the rows and $\sim75\%$ of the columns, while \textit{Mr. Clean - Greedy} kept $\sim95\%$ of the rows and $\sim30\%$ of the columns.

\subsection{RowCol IP vs. RowCol LP}
\begin{figure}[htbp]
    \centering
    \includegraphics[width=0.75\linewidth]{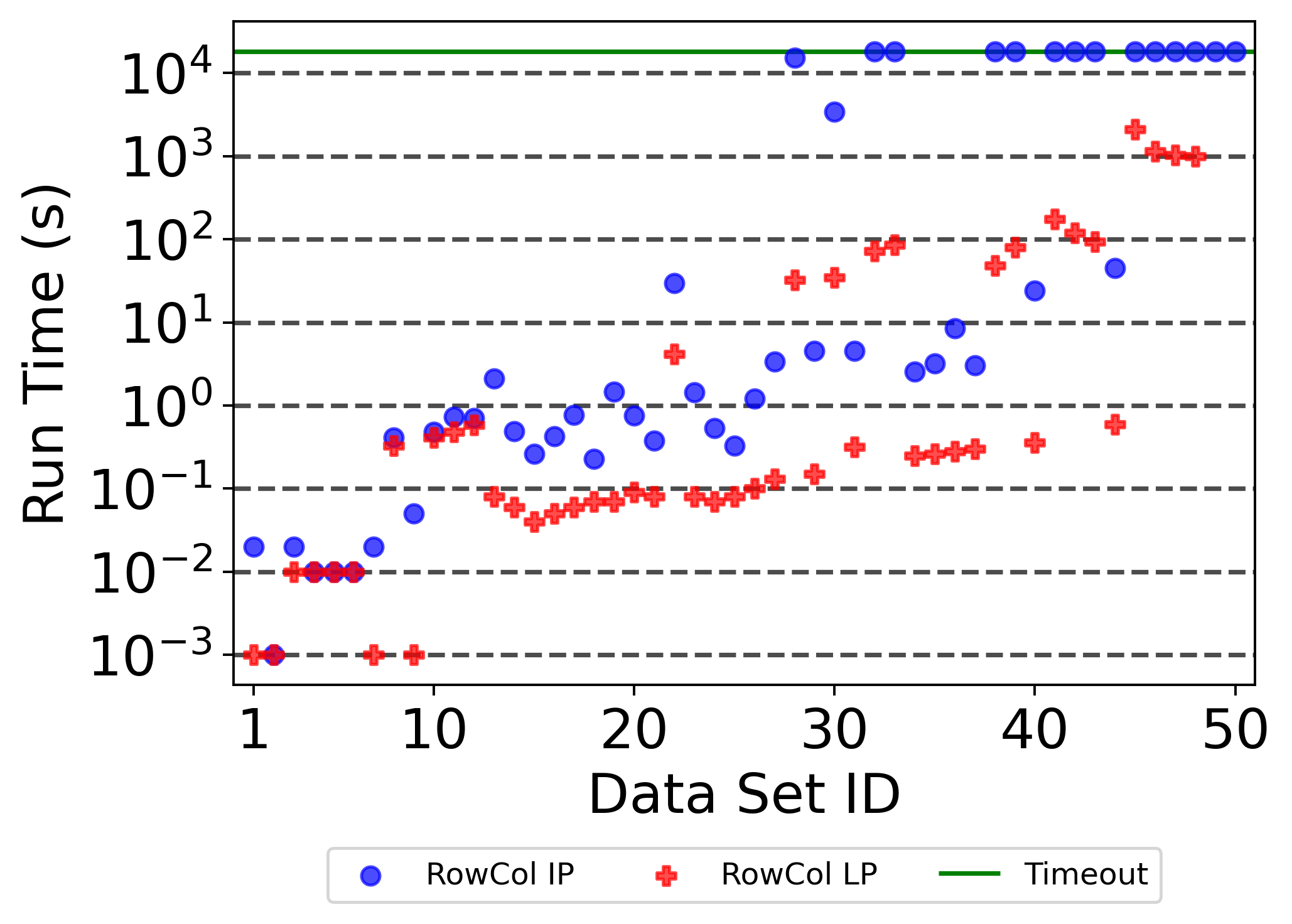}
    \caption{Run Time of RowCol IP and RowCol LP. The green line indicates the maximum allowed run time of 18000 seconds.}
    \label{fig:rowcol_runtime}
\end{figure}

In Figure \ref{fig:rowcol_runtime}, we compare run time of both \textit{RowCol} algorithms. On data sets 49 and 50, the \textit{RowCol IP} timed out and the \textit{RowCol LP} ran out of memory before solving. Overall, a major decrease in rum time (1-2 orders of magnitude) was achieved by the \textit{RowCol LP}, which as able to solve data sets 1-48 in under 2100 seconds each, with most problems solving in under 1 second. When both algorithms converged, they returned a matrix with the same dimension. Due to its ability to converge quicker and on a wider range of data sets, the \textit{RowCol LP} was selected for comparison to other programs in the $\gamma=0.0$ scenarios.

\subsection{MaxCol IP vs. Element IP}

\begin{figure}[htbp]
    \centering
    \includegraphics[width=0.75\linewidth]{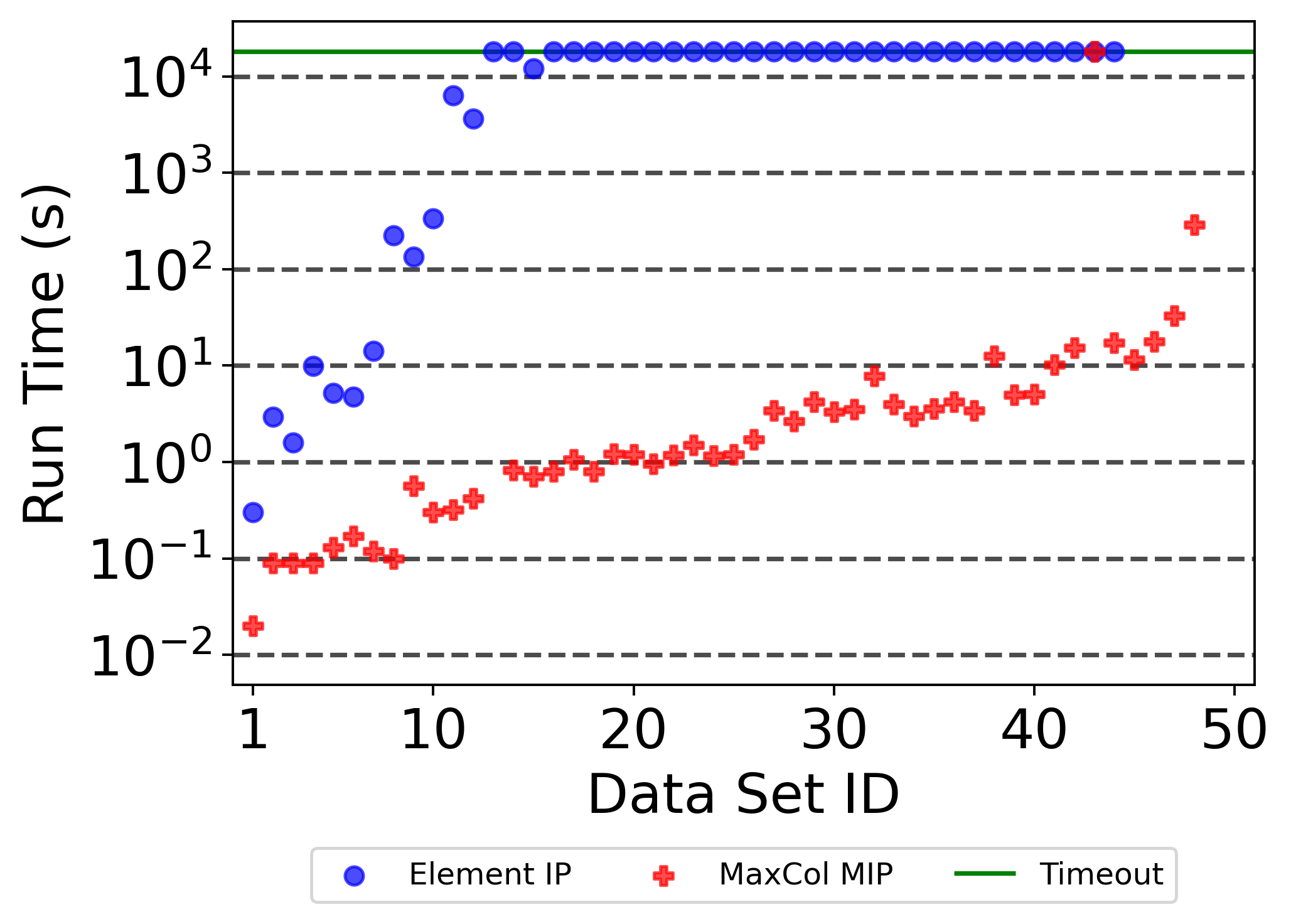}
    \caption{Run Time of MaxCol MIP and Element IP.  The green line indicates the maximum allowed run time of 18000 seconds.}
    \label{fig:maxcol_element_runtime}
\end{figure}

Next we compare the run times of \textit{MaxCol MIP} and \textit{Element IP} in Figure \ref{fig:maxcol_element_runtime}. The \textit{MaxCol MIP} was able to solve majority of the data sets within 10 seconds. However, data sets 13, 49, and 50 exhausted the available memory of the program, while \textit{MaxCol MIP} timed out on data set 43. On the other hand, \textit{Element IP} timed out for the majority of the data sets, only converging to a solution on data sets 1-12, and 14. Due to its reduced run time and its ability to converge for substantially more data sets, the \textit{MaxCol IP} was selected to run in the $\gamma=0.0$ scenarios.

\subsection{Comparing to Other Deletion Algorithms}
In this section, we evaluate the \textit{NoMiss} and \textit{Mr. Clean} algorithms against existing data cleaning routines. Each of the 50 data sets were cleaned to 3 different maximum missingness values: $\gamma=0.0$, $\gamma=0.05$, and $\gamma=0.1$.

\begin{figure}[htbp]
    \centering
    \subfloat[\label{fig:gamm0_elements}]{
        \includegraphics[width=0.75\linewidth]{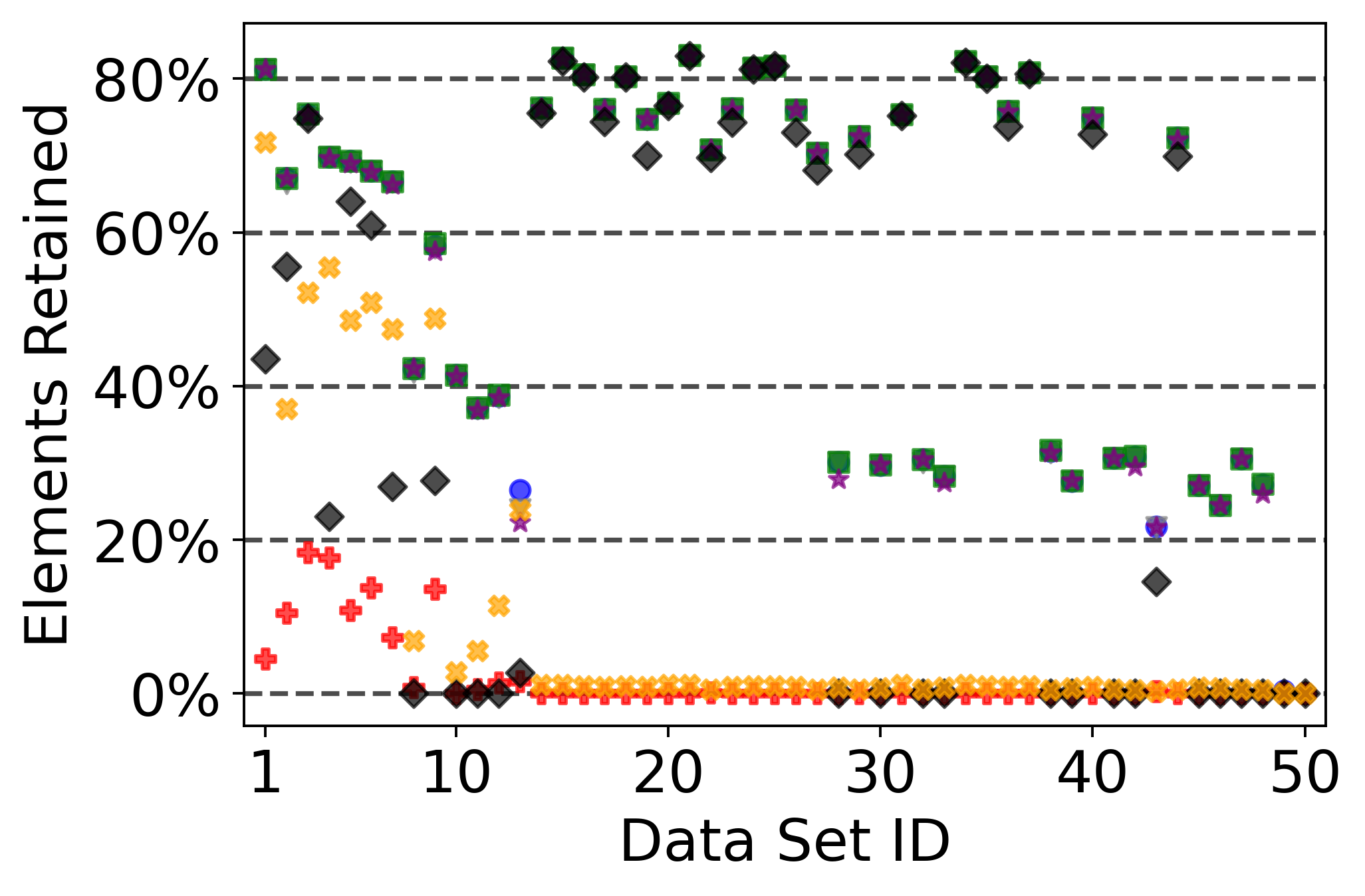}
    }
    \hfill
    \subfloat[\label{fig:gamm0_runtime}]{
        \includegraphics[width=0.75\linewidth]{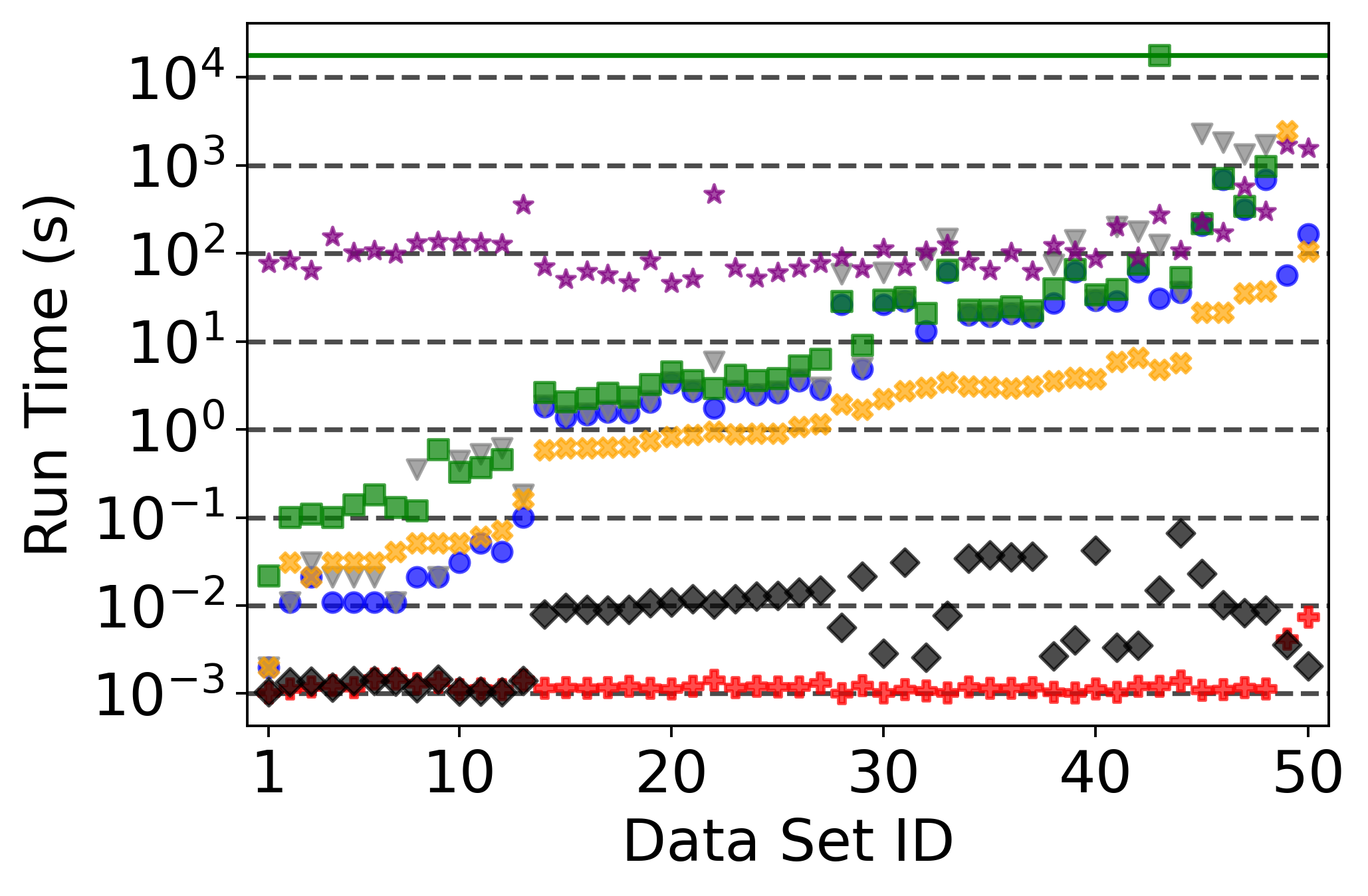}
    }
    \hfill
    \subfloat[\label{fig:gamm0_rows}]{
        \includegraphics[width=0.75\linewidth]{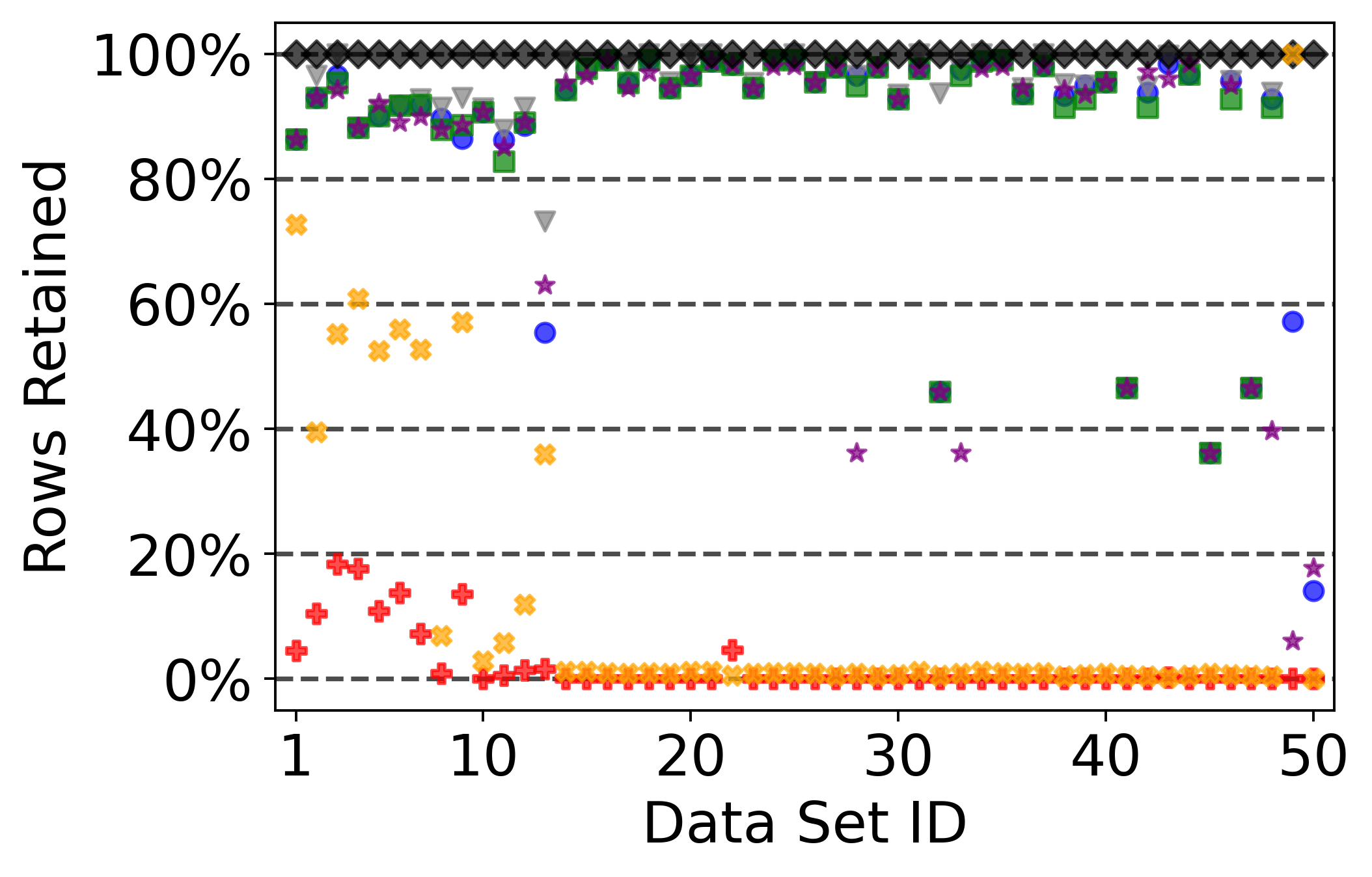}
    }
    \hfill
    \subfloat[\label{fig:gamm0_cols}]{
        \includegraphics[width=0.75\linewidth]{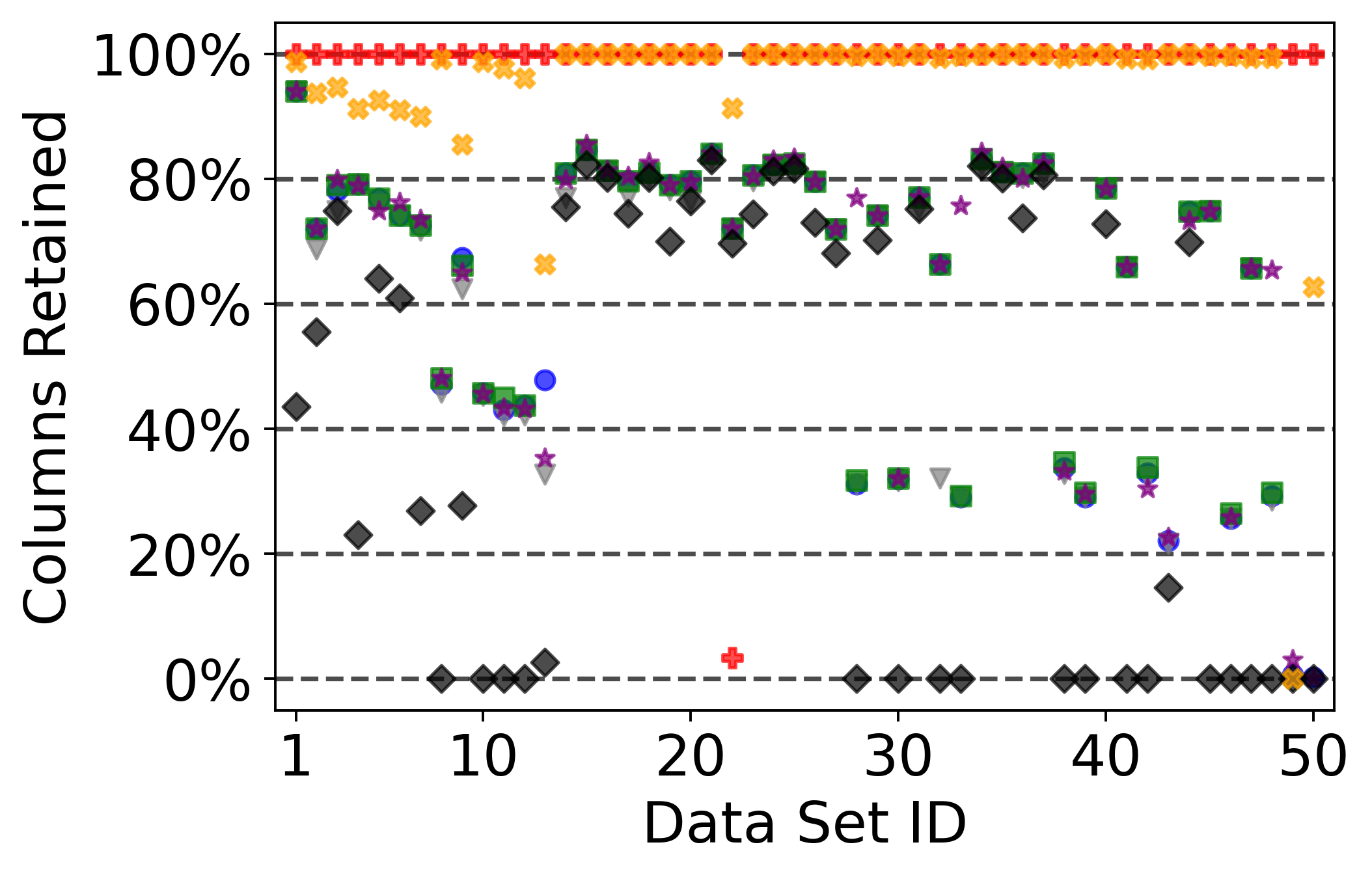}
    }
    \hfill
    \subfloat[\label{fig:gamma0_legends}]{
        \includegraphics[width=1\linewidth]{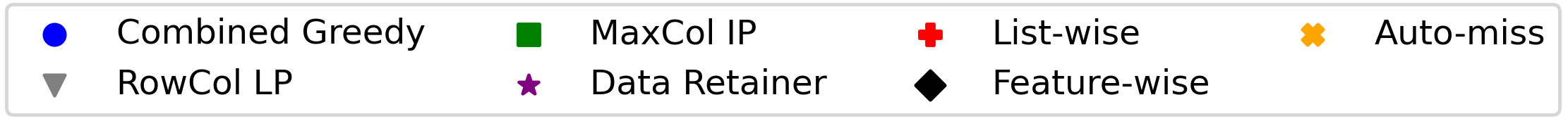}
    }
    \caption{Comparison of Deletion Algorithms at $\gamma=0.0$. \textbf{(a)} Percentage of Original Valid Elements Retained \textbf{(b)} Run Time. \textbf{(c)} Percentage of Original Rows Retained. \textbf{(d)} Percent of Original Columns Retained.} 
    \label{fig:gamma0}
\end{figure}

In the $\gamma=0.0$ scenario, we compared the combined \textit{greedy} algorithm, \textit{RowCol LP}, \textit{MaxCol MIP}, \textit{list-wise}, \textit{feature-wise}, \textit{auto-miss}, and \textit{DataRetainer} programs, as seen in Figure \ref{fig:gamma0}. Figure \ref{fig:gamm0_elements} shows the percentage of original valid element retained by each algorithm. When \textit{MaxCol MIP} solved the problem (data sets 1-12, 14-42, and 44-48), it achieved the most valid elements, as expected. Although, multiple algorithms often kept the maximum number of elements. For the remaining data sets, the maximum number of elements was obtained by \textit{greedy} in 3 scenarios and \textit{DataRetainer} in 1 scenario. The combined \textit{greedy} algorithm kept the maximum number of elements in 36 scenarios, 33 of which were proven optimal. All of the additional scenarios were within 0.5\% of maximum. The \textit{RowCol LP} performed slightly worse, keeping the optimal number of elements in only 15 of the 48 scenarios, while failing to converge for data sets 49 and 50. Of the sub-optimal solutions, all were withing 0.5\% of maximum, except for data set 13 which was $\sim$ 2.5\% from optimal. The \textit{DataRetainer} algorithms was the next most consistent, keeping the maximum number of elements in 15 scenarios. The \textit{Auto-miss} program never retained the maximum number of elements, removing more than 95\% of the possible elements in 38 scenarios. Since each row and/or column contained some missing data in several data sets, the \textit{list-wise} algorithm performed the worse overall, removing all elements in 36 scenarios. The \textit{feature-wise} algorithm removed all elements in 18 scenarios, but then performed well in 15 other scenarios, staying within 1\% of optimal.

\textit{DataRetainer} dominated the run time for the first 40 scenarios and then gave way to the \textit{MaxCol MIP} and \textit{RowCol LP}, as shown in Figure \ref{fig:gamm0_runtime}. In 27 scenarios, \textit{MaxCol MIP} ran in under 10 seconds. In fact, \textit{MaxCol IP}, \textit{RowCol LP}, and the \textit{combined greedy} ran at similar speeds for the first 27 scenarios. In Figure \ref{fig:gamm0_rows} we can see that \textit{list-wise} and \textit{auto-miss} removed the majority of the rows starting at data set 14. Similarly, \textit{feature-wise} deleted the most columns as seen in Figure \ref{fig:gamm0_cols}. In some scenarios, \textit{DataRetainer} created matrices much different than the \textit{NoMiss} algorithms. While keeping the number of elements within 1\% of optimal, the \textit{DataRetainer} solution to data set 33 contained 1/3 the rows and 2.6 times the number of columns as the greedy solution. This demonstrates that dramatically different solutions can have similar objective values.

\begin{figure}[htbp]
    \centering
    \subfloat[\label{fig:gamma5_elements}]{
        \includegraphics[width=0.75\linewidth]{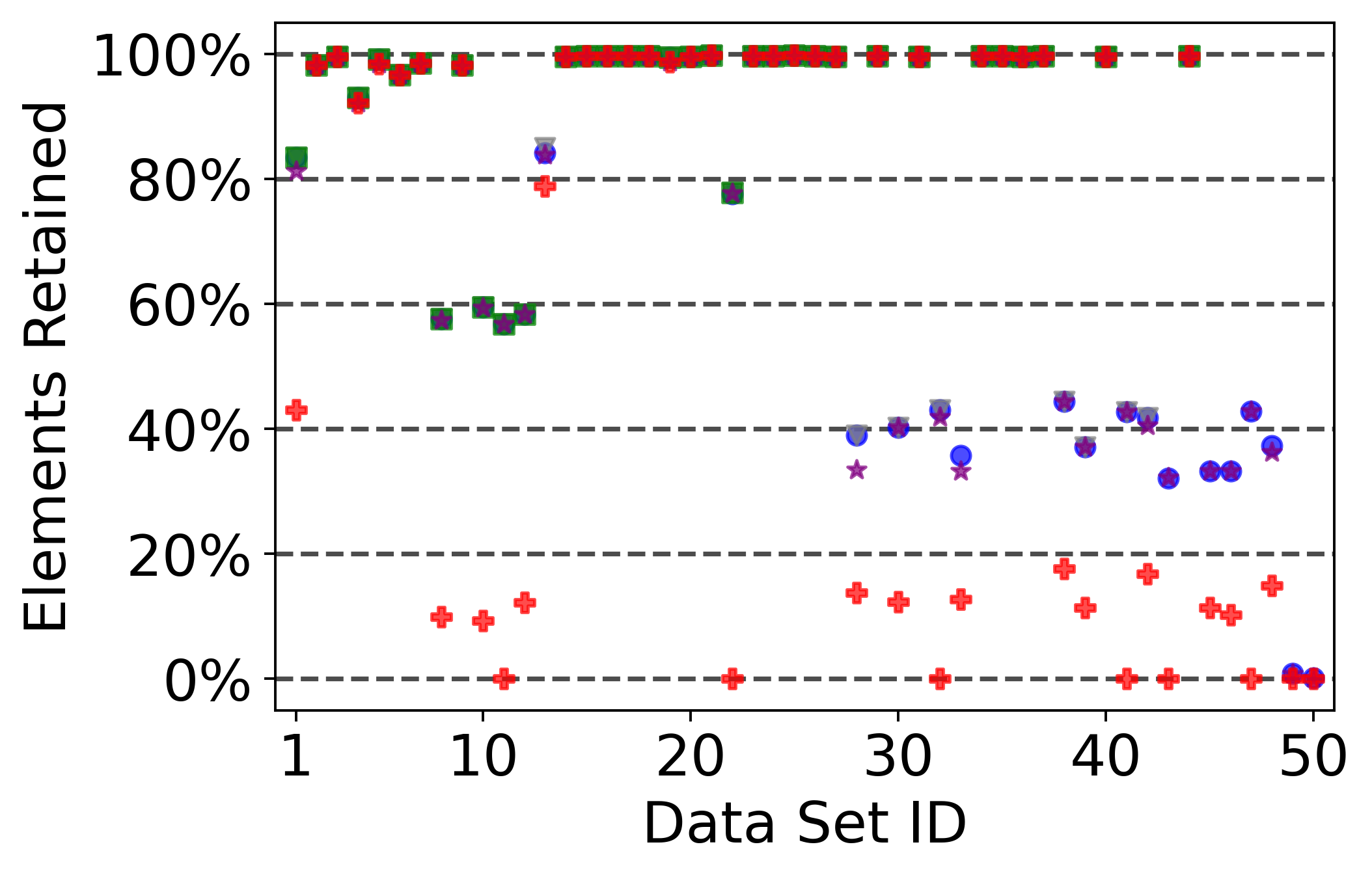}
    }
    \hfill
    \subfloat[\label{fig:gamma5_runtime}]{
        \includegraphics[width=0.75\linewidth]{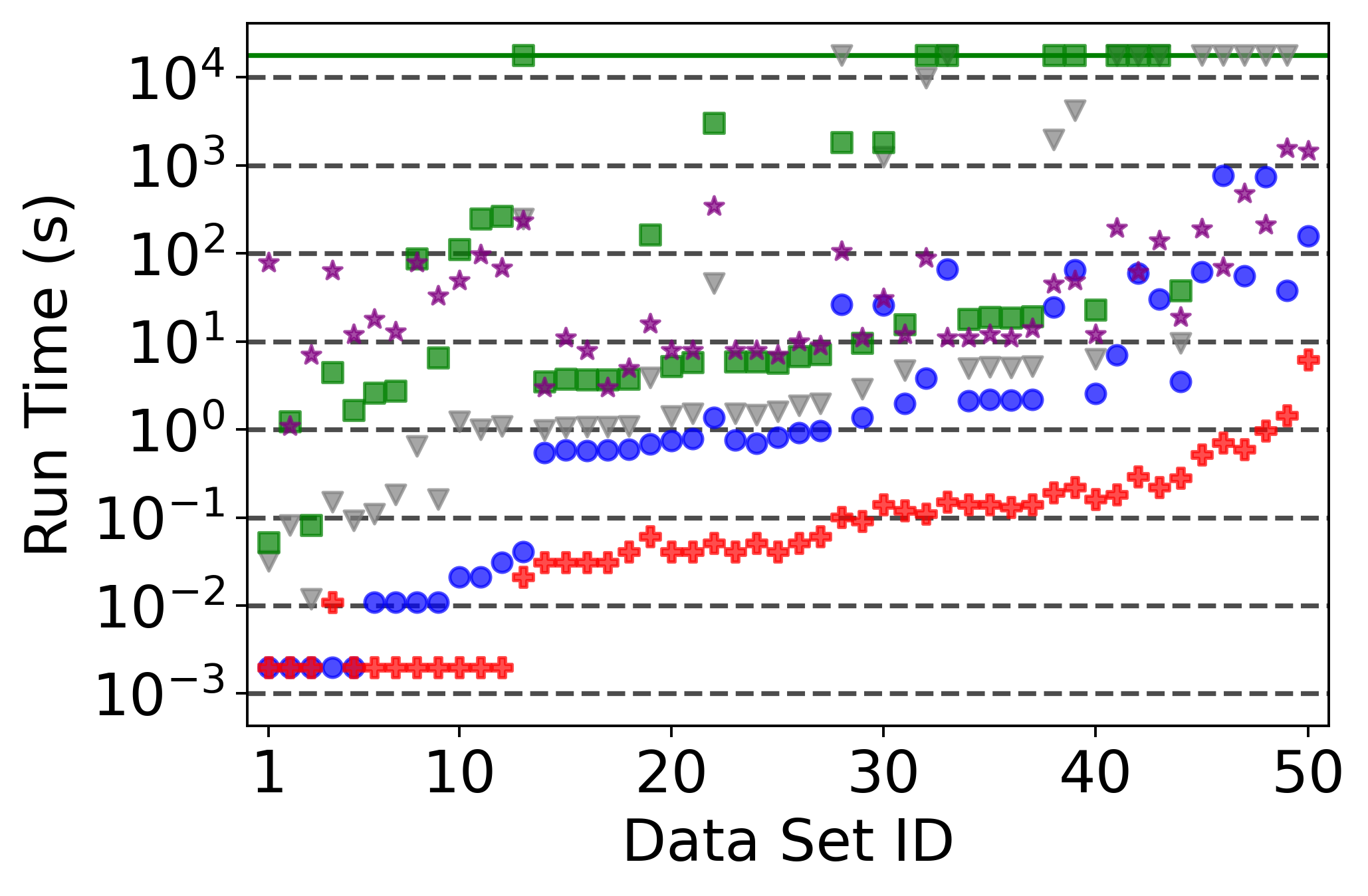}
    }
    \hfill
    \subfloat[\label{fig:gamma5_rows}]{
        \includegraphics[width=0.75\linewidth]{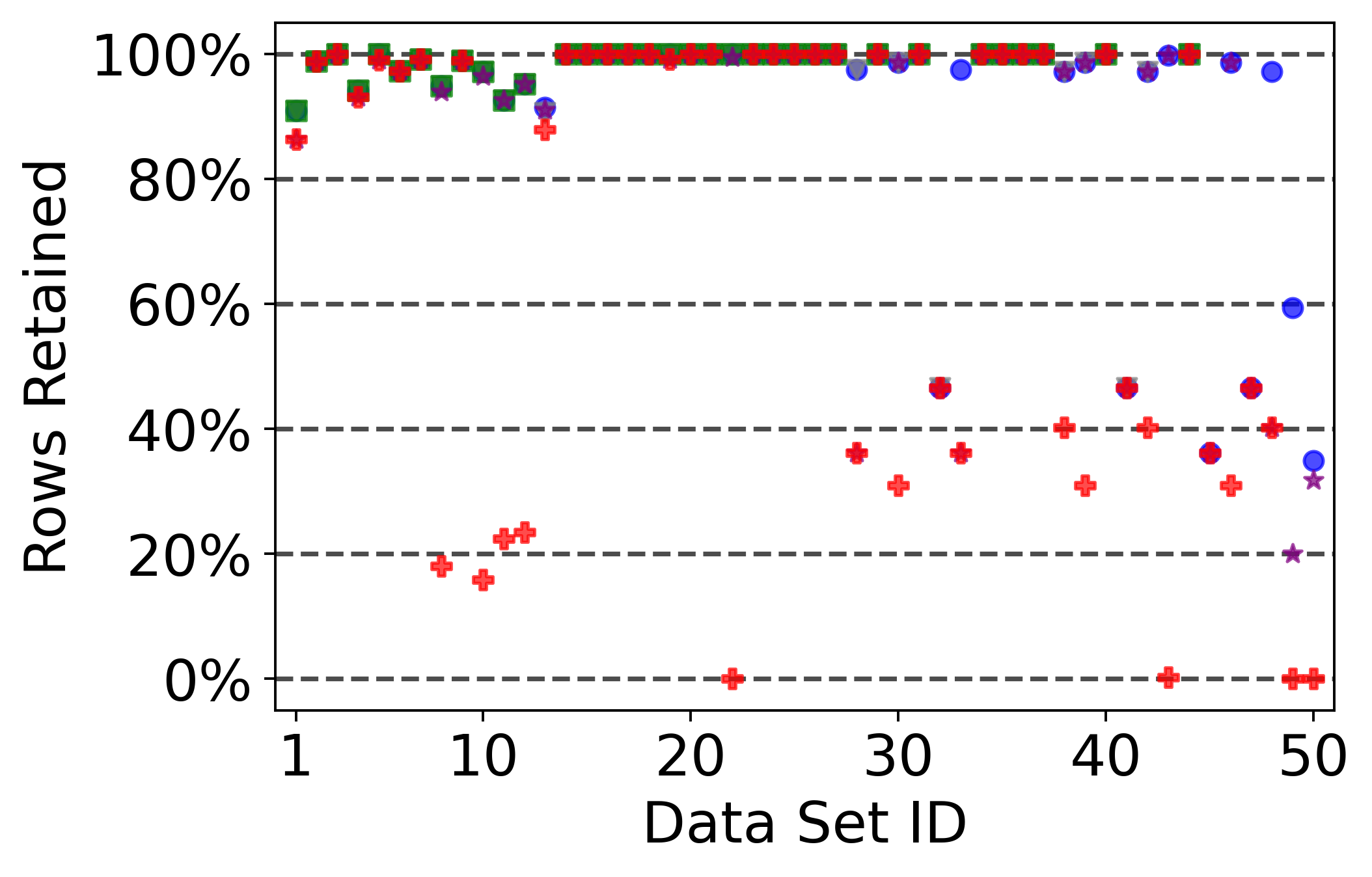}
    }
    \hfill
    \subfloat[\label{fig:gamma5_cols}]{
        \includegraphics[width=0.75\linewidth]{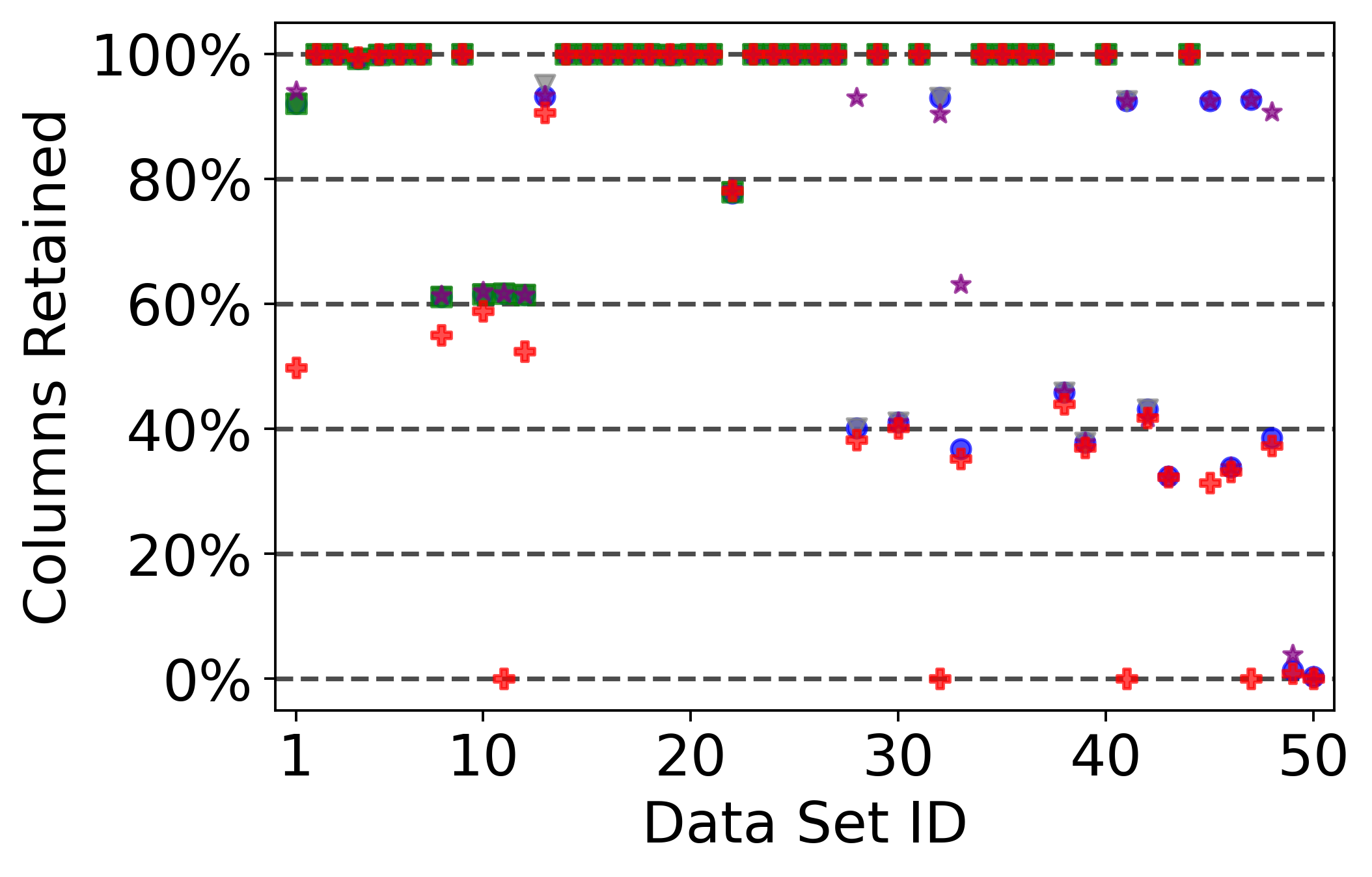}
    }
    \hfill
    \subfloat[\label{fig:gamma5_legends}]{
        \includegraphics[width=1\linewidth]{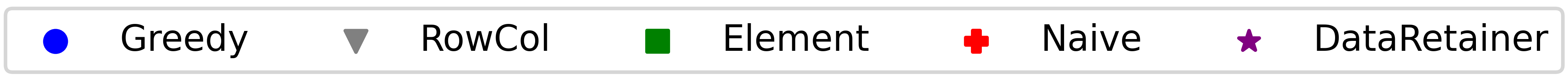}
    }
    \caption{Comparison of Deletion Algorithms at $\gamma=0.05$. \textbf{(a)} Percentage of Original Valid Elements Retained \textbf{(b)} Run Time. \textbf{(c)} Percentage of Original Rows Retained. \textbf{(d)} Percent of Original Columns Retained.} 
    \label{fig:gamma5}
\end{figure}

In the $\gamma=0.05$ and $\gamma=0.1$ scenarios, we compared the \textit{Mr. Clean} algorithms to \textit{naive} deletion and \textit{DataRetainer}. The \textit{naive} approach removes any row or column that exceeds the maximum percentage of missing data. Figure \ref{fig:gamma5} shows the results of the programs at $\gamma=0.05$. The \textit{Element IP} produced no solution for 16 data sets, 10 due to timeout and 6 due to memory limits. The remaining experiments were solved optimally. While the larger sets tended to be most difficult, data set 13 also plagued the program. The \textit{RowCol IP} optimally solved 41 of the data sets and produced no solution for 8 data sets. The only set which \textit{RowCol IP} solved sub-optimally was data set 10, in which it was within  0.002\% of optimal. \textit{Mr. Clean greedy} was able to clean all data sets, keeping the maximum number of valid elements in 42 sets and staying within 1\% of the maximum kept in the rest. These sub-optimal solutions tended to occur within the smaller data sets. \textit{DataRetainer} performed slightly worse, identifying the best solutions in 34 data sets. The sub-optimal solutions were also worse than \textit{Mr. Clean greedy}, missing over 2\% of the elements in three of the experiments. As expected, the \textit{naive} approach removed the most elements. In some cases no valid elements were kept. This occurred when all of the rows, or all of the columns, contained more than 5\% missing data.

The run times for $\gamma=0.05$ are compared in Figure \ref{fig:gamma5_runtime}. \textit{Element IP} and \textit{RowCol IP} timed out on 9 data sets and 10 data sets, respectively, while solving the majority of the small and medium size data sets in under 60 seconds. All data sets were cleaned by \textit{Mr. Clean greedy} in under 13 minutes, with most taking less than 1 minute. \textit{DataRetainer} fell in between the IPs and \textit{greedy} in terms of run time. Most data sets were cleaned within a minute, but the largest set took almost 27 minutes to clean. The \textit{naive} approach processed all of the files the quickest. Most data sets took less than 0.5 seconds, with the longest run time of 6 seconds occurring for the largest data set.

The \textit{DataRetainer} and \textit{Mr. clean} algorithms produced solutions with similar dimensions for majority of data sets. Notable exceptions are trails 28, 33, 48, and 49. Some of these differences resulted in $\sim5.5\%$ fewer elements (data set 28), while others resulted in only a small difference in elements ($1.1\%$ for data set 48). Instances of vastly different dimensions with a similar number of valid elements highlight the diversity of near optimal solutions. 

\begin{figure}[htbp]
    \centering
    \subfloat[\label{fig:gamma10_elements}]{
        \includegraphics[width=0.75\linewidth]{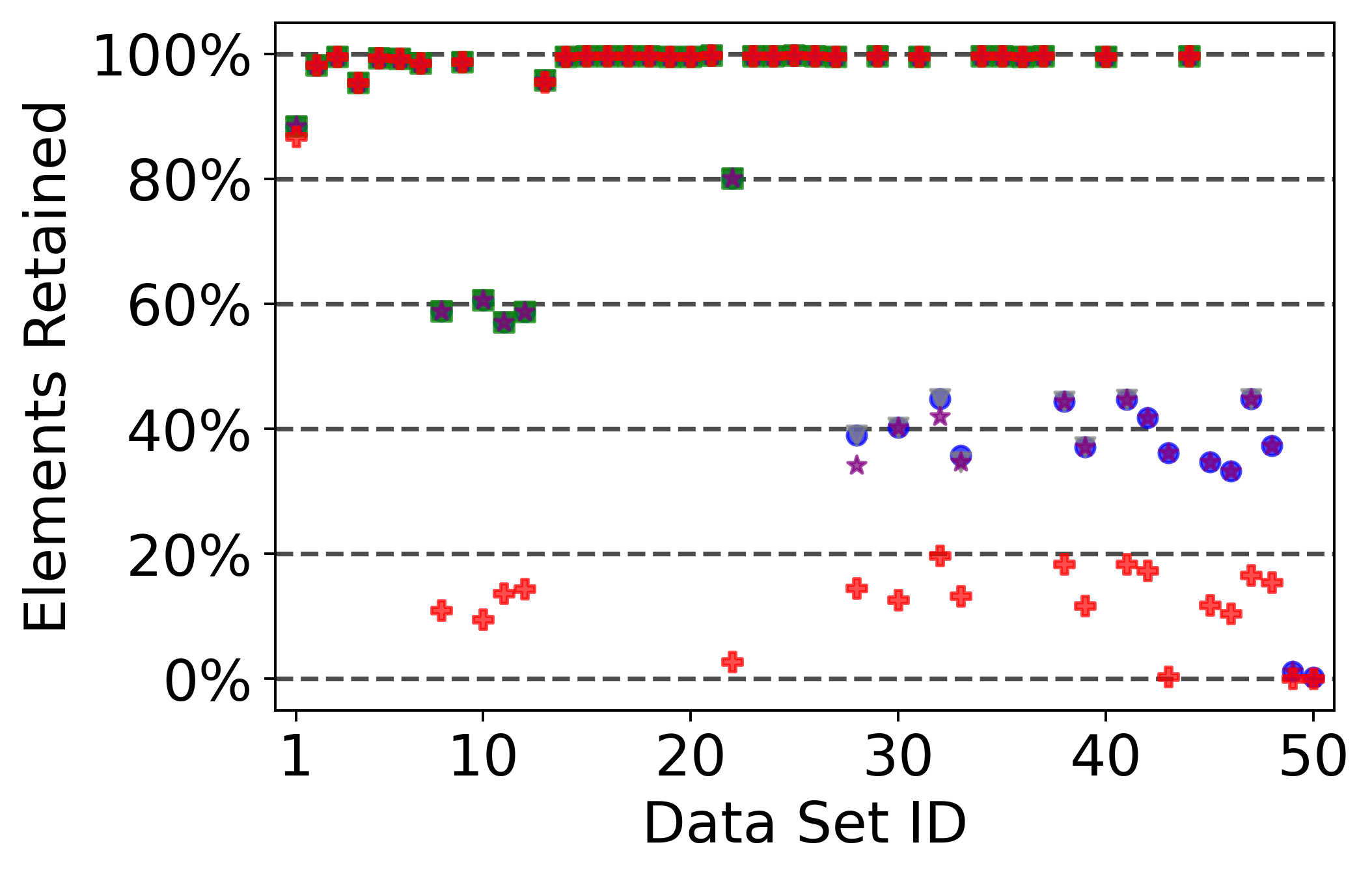}
    }
    \hfill
    \subfloat[\label{fig:gamma10_runtime}]{
        \includegraphics[width=0.75\linewidth]{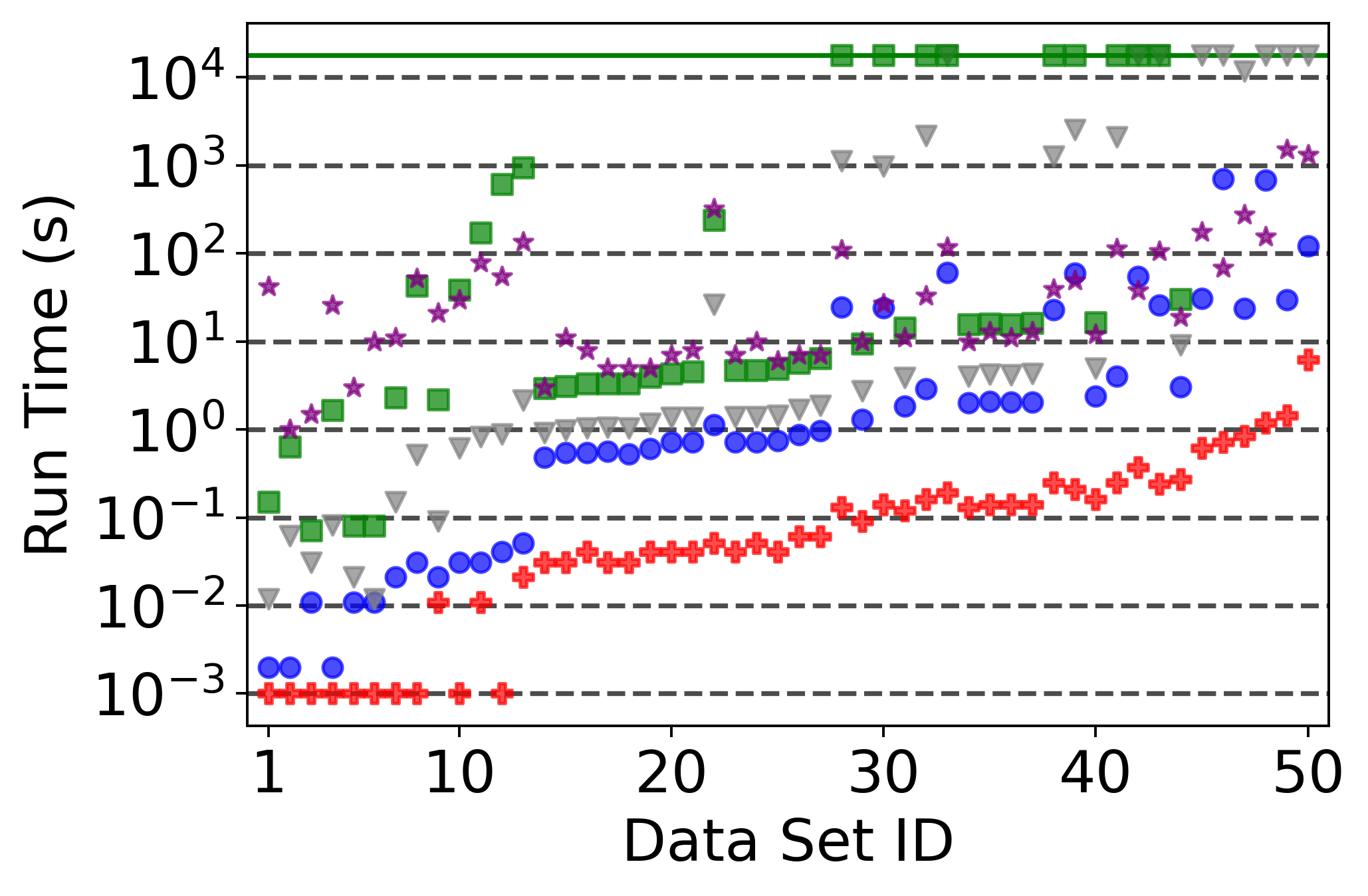}
    }
    \hfill
    \subfloat[\label{fig:gamma10_rows}]{
        \includegraphics[width=0.75\linewidth]{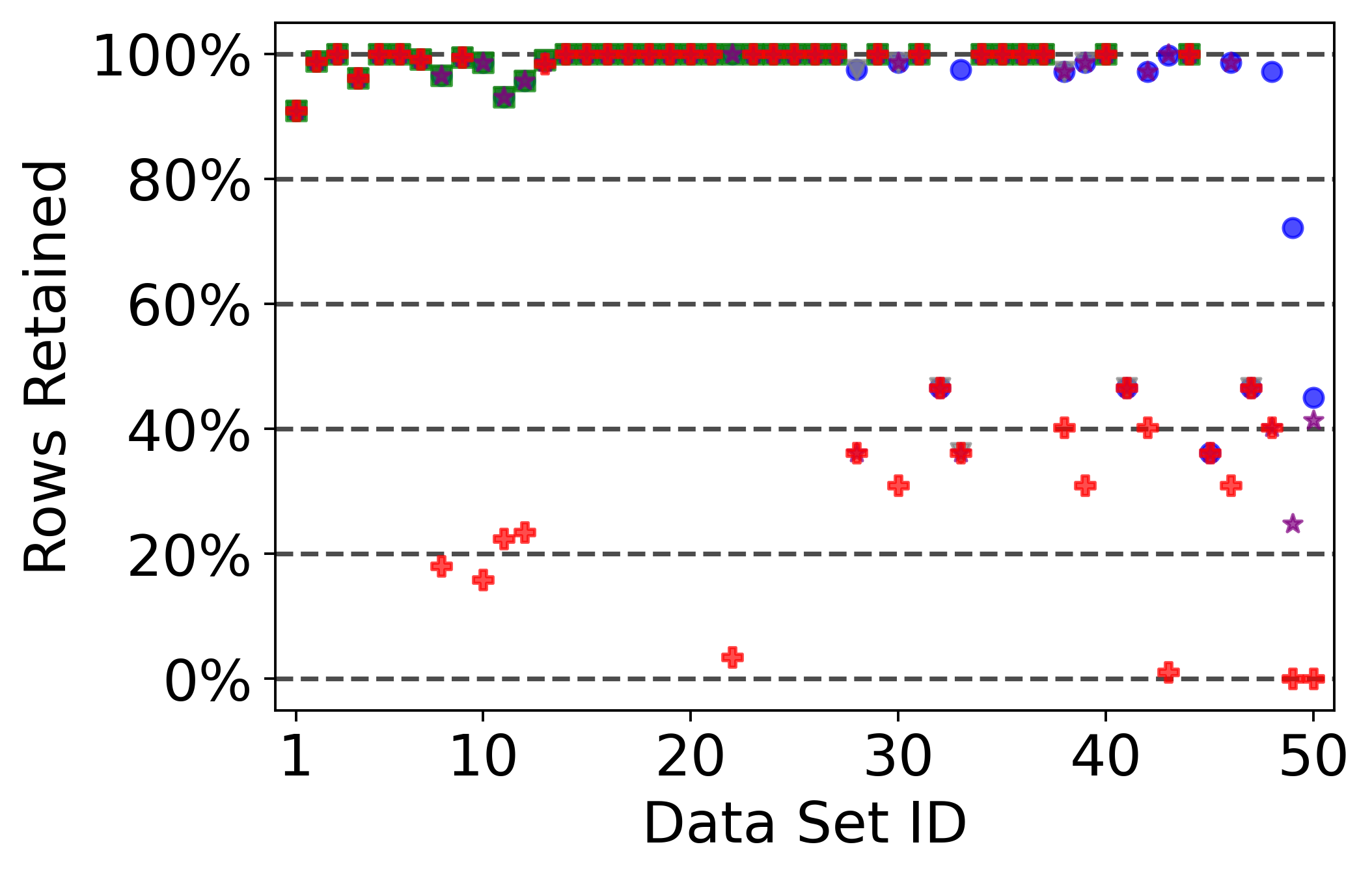}
    }
    \hfill
    \subfloat[\label{fig:gamma10_cols}]{
        \includegraphics[width=0.75\linewidth]{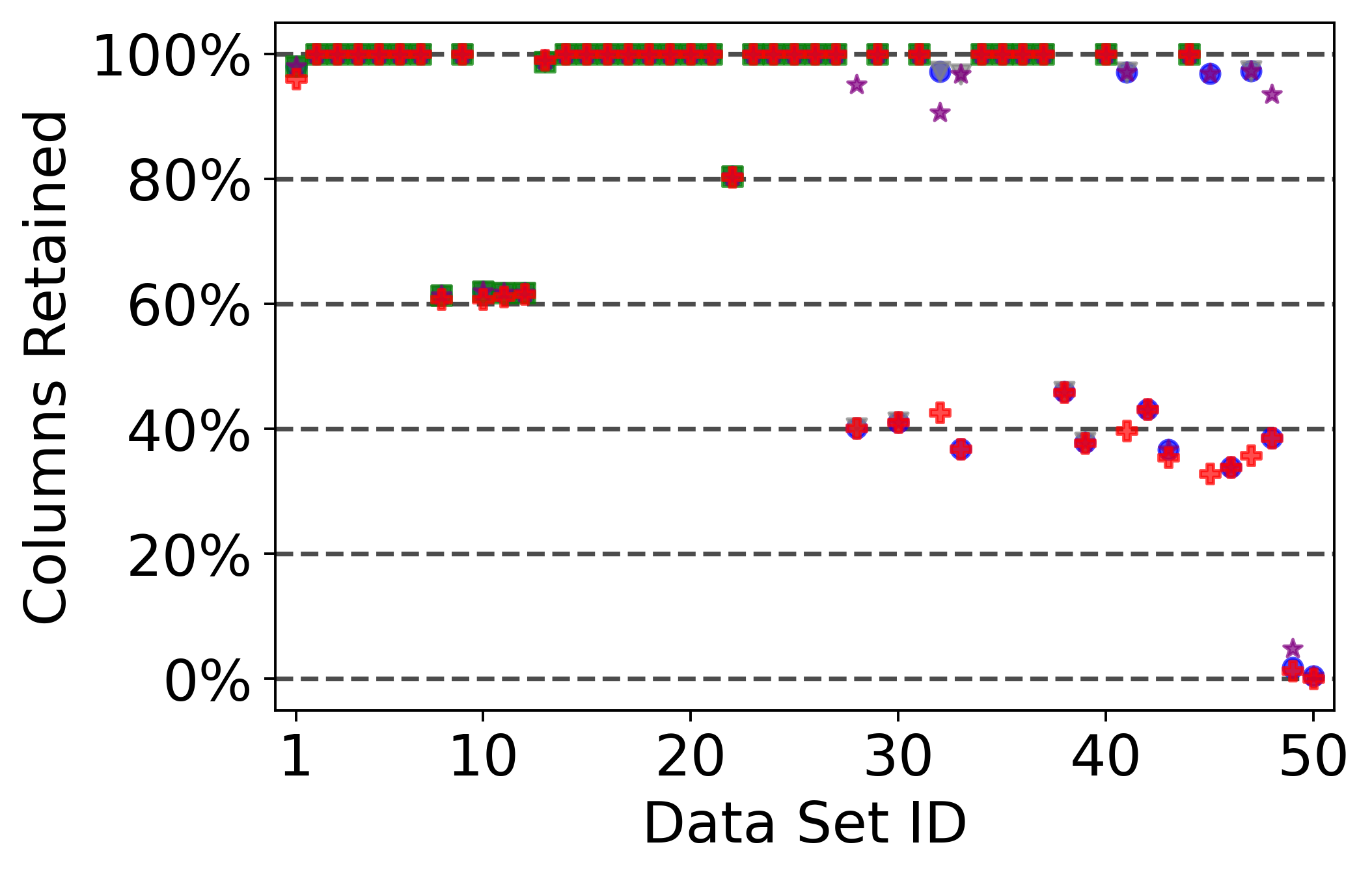}
    }
    \hfill
    \subfloat[\label{fig:gamma10_legends}]{
        \includegraphics[width=1\linewidth]{gamma10_legend.png}
    }
    \caption{Comparison of Deletion Algorithms at $\gamma=0.1$. \textbf{(a)} Percentage of Original Valid Elements Retained \textbf{(b)} Run Time. \textbf{(c)} Percentage of Original Rows Retained. \textbf{(d)} Percent of Original Columns Retained.} 
    \label{fig:gamma10}
\end{figure}

Figure \ref{fig:gamma10} shows the comparison between the deletion algorithms at $\gamma=0.1$. \textit{Element IP} solved 35 data sets optimally and returned no solution for the remaining 15. All of these data sets were difficult for the program at $\gamma=0.05$ as well. \textit{RowCol IP} failed to find a solution in 7 experiments, but retained the maximum number of elements in 42 experiments, 36 of which were proven optimal. Data set 33 proved the most difficult for \textit{RowCol IP}, where it kept 1\% fewer elements than \textit{greedy}. The \textit{Mr. Clean greedy} algorithm performed well again, retaining the most elements in 48 experiments. In the two cases where it failed to fine the best solution, it was still within 0.1\%. \textit{DataRetainer} performed slightly worse than \textit{greedy}, retaining the maximum number of elements in 43 experiments. However, in 3 of the experiments, \textit{DataRetainer} deleted over 1\% more elements than required. The \textit{naive} approach performed really well on some data sets, and then terribly on others.

The run times for $\gamma=0.1$ is shown in Figure \ref{fig:gamma10_runtime}. The \textit{Element IP} and \textit{DataRetainer} alternate longest run time. Towards the middle of the graph (data sets 14-31) most data sets contained less than 10\% missing data in each row and column. The run time for these experiments shows the difference in set-up and constraint checking time for the various methods. The \textit{RowCol IP} and \textit{greedy} algorithms show a similar run time pattern for many experiments, with the exception of the time out and near timeout runs. The solution dimensions between 4 of the algorithms were very consistent, as seen in Figures \ref{fig:gamma10_rows} and \ref{fig:gamma10_cols}. Decreases in the number of elements retained by the \textit{naive} algorithm are matched by decreases in either Figure \ref{fig:gamma10_rows} or \ref{fig:gamma10_cols}. Similar to other experiments, there were a few data sets where \textit{DataRetainer} found dramatically different solutions with a similar number valid elements compared to the \textit{Mr. Clean} algorithms. 

\section{Conclusion} \label{{sec:conclusion}}
In this manuscript we recast the \textit{Mr. Clean} IPs for the $\gamma=0.0$ scenario. We showed that the \textit{RowCol IP} can be solved as a LP, reducing runtime and increasing the file size that could be processed. For the \textit{Element IP}, we demonstrated that by partitioning the solution space and solving several simple MIPs in parallel, the run time could be dramatically reduced while increasing the maximum problem size. Finally, we added a new \textit{NoMiss greedy} algorithm, which supplements the \textit{Mr. Clean greedy} algorithm.

After comparing the \textit{NoMiss} programs to their \textit{Mr. Clean} counterparts, we evaluated both sets of programs against existing deletion algorithms. The new combined \textit{greedy} algorithm proved to be the best balance between run time and retaining elements over all experiments. It was able to solve all problems and retained the most elements in 127 of 150 scenarios. The $\gamma=0.0$ experiments gave the algorithm the most difficulty. However, many of these experiments were optimally solved by the \textit{MaxCol MIP}, which ran sufficiently fast to use in place of \textit{greedy} when $\gamma = 0.0$. This MIP handled all but 4 of data files in our 5 hour limit, with most solving within 10 seconds. The \textit{RowCol} algorithms showed a slight improvement over \textit{greedy} in a few scenarios, but only in ones that were successfully cleaned by \textit{Element IP}. The \textit{list-wise}, \textit{feature-wise}, \textit{naive}, and \textit{auto-miss} algorithms while running very fast, removed considerably more elements than other approaches. The \textit{DataRetainer} performed well on most of the scenarios, but required constant user interaction for each data set. We therefore recommend using the \textit{MaxCol MIP} when cleaning data for $\gamma=0.0$ and utilizing the \textit{combined greedy} algorithm for other $\gamma$ value, or if the data set is too large for the \textit{MaxCol MIP}. Finally, the \textit{Element IP} can be employed on small or medium data sets when extra runtime and memory are available.

Possible future work in this area could include creating a distributed version of \textit{Element IP} for $\gamma > 0.0$ scenarios or rerunning the IP/MIPs while controlling the minimum number of remaining rows and columns. It is also worth investigating the downstream impacts of partial deletion to imputation, along with better understanding the vast differences in matrix dimensions for near optimal solutions.

\printbibliography

@article{Climer2014,
   author = {Sharlee Climer and Alan R. Templeton and Weixiong Zhang},
   doi = {10.1371/journal.pcbi.1003766},
   isbn = {2014070504},
   issn = {15537358},
   issue = {9},
   journal = {PLoS Computational Biology},
   pmid = {25233071},
   title = {Allele-Specific Network Reveals Combinatorial Interaction That Transcends Small Effects in Psoriasis GWAS},
   volume = {10},
   year = {2014},
}

@article{Li2014,
   author = {Bingshan Li and Lam C. Tsoi and William R. Swindell and Johann E. Gudjonsson and Trilokraj Tejasvi and Andrew Johnston and Jun Ding and Philip E. Stuart and Xianying Xing and James J. Kochkodan and John J. Voorhees and Hyun M. Kang and Rajan P. Nair and Goncalo R. Abecasis and James T. Elder},
   doi = {10.1038/jid.2014.28},
   issn = {15231747},
   issue = {7},
   journal = {Journal of Investigative Dermatology},
   pages = {1828-1838},
   pmid = {24441097},
   publisher = {Elsevier Masson SAS},
   title = {Transcriptome analysis of psoriasis in a large case-control sample: RNA-seq provides insights into disease mechanisms},
   volume = {134},
   url = {http://dx.doi.org/10.1038/jid.2014.28},
   year = {2014},
}

@article{NewmanD2014,
   author = {Daniel A. Newman},
   doi = {10.1177/1094428114548590},
   issn = {15527425},
   issue = {4},
   journal = {Organizational Research Methods},
   month = {10},
   pages = {372-411},
   publisher = {SAGE Publications Inc.},
   title = {Missing Data: Five Practical Guidelines},
   volume = {17},
   year = {2014},
}

@article{WoodsAD2021,
   author = {Adrienne D Woods and Pam Davis-Kean and Max Halvorson and Kevin King and Jessica R Logan and Menglin Xu and Sierra Bainter and Denver Brown and James M Clay and Rick A Cruz and Mahmoud M Elsherif and Daria Gerasimova and Keven Joyal-Desmarais and David Moreau and Jayson Nissen and Kathleen Schmidt and Alex Uzdavines and Ben Van Dusen and Martin Vasilev},
   doi = {10.31234/osf.io/mdw5r},
   journal = {PsyArXiv},
   title = {Missing Data and Multiple Imputation Decision Tree},
   url = {https://doi.org/10.31234/osf.io/mdw5r},
   year = {2021},
}

@article{Souto2015,
   author = {Marcilio C.P.D. Souto and Pablo A. Jaskowiak and Ivan G. Costa},
   doi = {10.1186/s12859-015-0494-3},
   issn = {14712105},
   issue = {1},
   journal = {BMC Bioinformatics},
   month = {2},
   pmid = {25888091},
   publisher = {BioMed Central Ltd.},
   title = {Impact of missing data imputation methods on gene expression clustering and classification},
   volume = {16},
   year = {2015},
}

@article{ChoiJ2019,
   author = {Jungyeon Choi and Olaf M. Dekkers and Saskia le Cessie},
   doi = {10.1007/s10654-018-0447-z},
   issn = {15737284},
   issue = {1},
   journal = {European Journal of Epidemiology},
   month = {1},
   pages = {23-36},
   pmid = {30341708},
   publisher = {Springer Netherlands},
   title = {A comparison of different methods to handle missing data in the context of propensity score analysis},
   volume = {34},
   year = {2019},
}

@article{vanGinkel2020,
   author = {Joost R. van Ginkel and Marielle Linting and Ralph C.A. Rippe and Anja van der Voort},
   doi = {10.1080/00223891.2018.1530680},
   issn = {00223891},
   issue = {3},
   journal = {Journal of Personality Assessment},
   month = {5},
   pages = {297-308},
   pmid = {30657714},
   publisher = {Routledge},
   title = {Rebutting Existing Misconceptions About Multiple Imputation as a Method for Handling Missing Data},
   volume = {102},
   year = {2020},
}

@article{MadleyDowd2019,
   author = {Paul Madley-Dowd and Rachael Hughes and Kate Tilling and Jon Heron},
   doi = {10.1016/j.jclinepi.2019.02.016},
   issn = {18785921},
   journal = {Journal of Clinical Epidemiology},
   month = {6},
   pages = {63-73},
   pmid = {30878639},
   publisher = {Elsevier USA},
   title = {The proportion of missing data should not be used to guide decisions on multiple imputation},
   volume = {110},
   year = {2019},
}

@article{PeetersM2015,
   author = {Margot Peeters and Mariëlle Zondervan-Zwijnenburg and Gerko Vink and Rens van de Schoot},
   doi = {10.1080/17405629.2015.1049526},
   issn = {17405610},
   issue = {4},
   journal = {European Journal of Developmental Psychology},
   month = {7},
   pages = {377-394},
   publisher = {Psychology Press Ltd},
   title = {How to handle missing data: A comparison of different approaches},
   volume = {12},
   year = {2015},
}

@article{DubeyA2021,
   author = {Aditya Dubey and Akhtar Rasool},
   doi = {10.1038/s41598-021-03438-x},
   issn = {20452322},
   issue = {1},
   journal = {Scientific Reports},
   month = {12},
   pmid = {34934107},
   publisher = {Nature Research},
   title = {Efficient technique of microarray missing data imputation using clustering and weighted nearest neighbour},
   volume = {11},
   year = {2021},
}

@article{FarswanA2020,
   author = {Akanksha Farswan and Anubha Gupta and Ritu Gupta and Gurvinder Kaur},
   doi = {10.3389/fonc.2019.01442},
   issn = {2234943X},
   journal = {Frontiers in Oncology},
   month = {1},
   publisher = {Frontiers Media S.A.},
   title = {Imputation of Gene Expression Data in Blood Cancer and Its Significance in Inferring Biological Pathways},
   volume = {9},
   year = {2020},
}

@article{ClimerS2021,
   author = {Sharlee Climer},
   doi = {10.1016/j.patter.2021.100374},
   issn = {26663899},
   issue = {12},
   journal = {Patterns},
   month = {12},
   publisher = {Cell Press},
   title = {Connecting the dots: The boons and banes of network modeling},
   volume = {2},
   year = {2021},
}

@inproceedings{DayalB1994,
   author = {Bhupinder S Dayal and John F Macgregor and Paul A Taylor and R Kildaw and S Marcikic},
   title = {Application of feedforward: neural networks and partial least squares regression for modelling kappa number in a continuous Kamyr digester: how multivariate data analysis might help pulping},
   year = {1994},
}

@misc{DunnK2011,
   author = {Kevin Dunn},
   journal = {OpenMV.net},
   month = {10},
   title = {Kamyr digester-OpenMV.net Datasets.},
   url = {https://openmv.net/info/kamyr-digester},
   year = {2011},
}

@article{SzaboP2023,
   author = {Peter M Szabo and Amir Vajdi and Namit Kumar and Michael Y Tolstorukov and Benjamin J Chen and Robin Edwards and Keith L Ligon and Scott D Chasalow and Kin-Hoe Chow and Aniket Shetty and Mohan Bolisetty and James L Holloway and Ryan Golhar and Brian A Kidd and Philip Ansumana Hull and Jeff Houser and Logan Vlach and Nathan O Siemers and Saurabh Saha},
   doi = {10.1038/s41598-023-28480-9},
   isbn = {0123456789},
   issue = {3051},
   journal = {Scientific Reports},
   title = {Cancer-associated fibroblasts are the main contributors to epithelial-to-mesenchymal signatures in the tumor microenvironment},
   volume = {13},
   url = {https://doi.org/10.1038/s41598-023-28480-9},
   year = {2023},
}

@article{YangC2021,
   author = {Chengran Yang and Fabiana H G Farias and Laura Ibanez and Adam Suhy and Brooke Sadler and Maria Victoria Fernandez and Fengxian Wang and Joseph L Bradley and Brett Eiffert and Jorge A Bahena and John P Budde and Zeran Li and Umber Dube and Yun Ju Sung and Kathie A Mihindukulasuriya and John C Morris and Anne M Fagan and Richard J Perrin and Bruno A Benitez and Herve Rhinn and Oscar Harari and Carlos Cruchaga},
   doi = {10.1038/s41593-021-00886-6},
   journal = {Nature Neuroscience},
   month = {9},
   pages = {1302-1312},
   title = {Genomic atlas of the proteome from brain, CSF and plasma prioritizes proteins implicated in neurological disorders},
   volume = {24},
   url = {https://doi.org/10.1038/s41593-021-00886-6},
   year = {2021},
}

@article{RubinD1975,
   author = {Donald B. Rubin},
   doi = {10.1002/J.2333-8504.1975.TB01053.X},
   issn = {2333-8504},
   issue = {1},
   journal = {ETS Research Bulletin Series},
   month = {6},
   pages = {i-19},
   publisher = {John Wiley & Sons, Ltd},
   title = {INFERENCE AND MISSING DATA},
   volume = {1975},
   url = {https://onlinelibrary.wiley.com/doi/full/10.1002/j.2333-8504.1975.tb01053.x},
   year = {1975},
}

@misc{Austin2021,
   author = {Peter C. Austin and Ian R. White and Douglas S. Lee and Stef van Buuren},
   doi = {10.1016/j.cjca.2020.11.010},
   issn = {0828282X},
   issue = {9},
   journal = {Canadian Journal of Cardiology},
   month = {9},
   pages = {1322-1331},
   pmid = {33276049},
   publisher = {Elsevier Inc.},
   title = {Missing Data in Clinical Research: A Tutorial on Multiple Imputation},
   volume = {37},
   year = {2021},
}

@article{MrClean,
   author = {Kenneth Smith and Sharlee Climer},
   title = {Mr. Clean: An Ensemble of Data Cleaning Algorithms for Increased Data Retention},
}

@article{Briggs2003,
   author = {Andrew Briggs and Taane, Clark and Jane Wolstenholme and Philip Clarke},
   doi = {10.1002/hec.766},
   issue = {5},
   journal = {Health Econcomics},
   pages = {377-392},
   pmid = {12720255},
   title = {Missing.... presumed at random: cost-analysis of incomplete data},
   volume = {12},
   url = {https://onlinelibrary.wiley.com/doi/10.1002/hec.766},
   year = {2003},
}

@article{Petrazzini2021,
   author = {Ben Omega Petrazzini and Hugo Naya and Fernando Lopez-Bello and Gustavo Vazquez and Lucía Spangenberg},
   doi = {10.1186/s13040-021-00274-7},
   issn = {17560381},
   issue = {1},
   journal = {BioData Mining},
   month = {12},
   publisher = {BioMed Central Ltd},
   title = {Evaluation of different approaches for missing data imputation on features associated to genomic data},
   volume = {14},
   year = {2021},
}

@article{HapMapIII,
   author = {Kelly A. Frazer and Dennis G. Ballinger and David R. Cox and David A. Hinds and Laura L. Stuve and Richard A. Gibbs and John W. Belmont and Andrew Boudreau and Paul Hardenbol and Suzanne M. Leal and Shiran Pasternak and David A. Wheeler and Thomas D. Willis and Fuli Yu and Huanming Yang and Changqing Zeng and Yang Gao and Haoran Hu and Weitao Hu and Chaohua Li and Wei Lin and Siqi Liu and Hao Pan and Xiaoli Tang and Jian Wang and Wei Wang and Jun Yu and Bo Zhang and Qingrun Zhang and Hongbin Zhao and Hui Zhao and Jun Zhou and Stacey B. Gabriel and Rachel Barry and Brendan Blumenstiel and Amy Camargo and Matthew Defelice and Maura Faggart and Mary Goyette and Supriya Gupta and Jamie Moore and Huy Nguyen and Robert C. Onofrio and Melissa Parkin and Jessica Roy and Erich Stahl and Ellen Winchester and Liuda Ziaugra and David Altshuler and Yan Shen and Zhijian Yao and Wei Huang and Xun Chu and Yungang He and Li Jin and Yangfan Liu and Yayun Shen and Weiwei Sun and Haifeng Wang and Yi Wang and Ying Wang and Xiaoyan Xiong and Liang Xu and Mary M.Y. Waye and Stephen K.W. Tsui and Hong Xue and J. Tze Fei Wong and Luana M. Galver and Jian Bing Fan and Kevin Gunderson and Sarah S. Murray and Arnold R. Oliphant and Mark S. Chee and Alexandre Montpetit and Fanny Chagnon and Vincent Ferretti and Martin Leboeuf and Jean François Olivier and Michael S. Phillips and Stéphanie Roumy and Clémentine Sallée and Andrei Verner and Thomas J. Hudson and Pui Yan Kwok and Dongmei Cai and Daniel C. Koboldt and Raymond D. Miller and Ludmila Pawlikowska and Patricia Taillon-Miller and Ming Xiao and Lap Chee Tsui and William Mak and Qiang Song You and Paul K.H. Tam and Yusuke Nakamura and Takahisa Kawaguchi and Takuya Kitamoto and Takashi Morizono and Atsushi Nagashima and Yozo Ohnishi and Akihiro Sekine and Toshihiro Tanaka and Tatsuhiko Tsunoda and Panos Deloukas and Christine P. Bird and Marcos Delgado and Emmanouil T. Dermitzakis and Rhian Gwilliam and Sarah Hunt and Jonathan Morrison and Don Powell and Barbara E. Stranger and Pamela Whittaker and David R. Bentley and Mark J. Daly and Paul I.W. De Bakker and Jeff Barrett and Yves R. Chretien and Julian Maller and Steve McCarroll and Nick Patterson and Itsik Pe'Er and Alkes Price and Shaun Purcell and Daniel J. Richter and Pardis Sabeti and Richa Saxena and Stephen F. Schaffner and Pak C. Sham and Patrick Varilly and Lincoln D. Stein and Lalitha Krishnan and Albert Vernon Smith and Marcela K. Tello-Ruiz and Gudmundur A. Thorisson and Aravinda Chakravarti and Peter E. Chen and David J. Cutler and Carl S. Kashuk and Shin Lin and Gonçalo R. Abecasis and Weihua Guan and Yun Li and Heather M. Munro and Zhaohui Steve Qin and Daryl J. Thomas and Gilean McVean and Adam Auton and Leonardo Bottolo and Niall Cardin and Susana Eyheramendy and Colin Freeman and Jonathan Marchini and Simon Myers and Chris Spencer and Matthew Stephens and Peter Donnelly and Lon R. Cardon and Geraldine Clarke and David M. Evans and Andrew P. Morris and Bruce S. Weir and Todd A. Johnson and James C. Mullikin and Stephen T. Sherry and Michael Feolo and Andrew Skol and Houcan Zhang and Ichiro Matsuda and Yoshimitsu Fukushima and Darryl R. MacEr and Eiko Suda and Charles N. Rotimi and Clement A. Adebamowo and Ike Ajayi and Toyin Aniagwu and Patricia A. Marshall and Chibuzor Nkwodimmah and Charmaine D.M. Royal and Mark F. Leppert and Missy Dixon and Andy Peiffer and Renzong Qiu and Alastair Kent and Kazuto Kato and Norio Niikawa and Isaac F. Adewole and Bartha M. Knoppers and Morris W. Foster and Ellen Wright Clayton and Jessica Watkin and Donna Muzny and Lynne Nazareth and Erica Sodergren and George M. Weinstock and Imtaz Yakub and Bruce W. Birren and Richard K. Wilson and Lucinda L. Fulton and Jane Rogers and John Burton and Nigel P. Carter and Christopher M. Clee and Mark Griffiths and Matthew C. Jones and Kirsten McLay and Robert W. Plumb and Mark T. Ross and Sarah K. Sims and David L. Willey and Zhu Chen and Hua Han and Le Kang and Martin Godbout and John C. Wallenburg and Paul L'Archevêque and Guy Bellemare and Koji Saeki and Hongguang Wang and Daochang An and Hongbo Fu and Qing Li and Zhen Wang and Renwu Wang and Arthur L. Holden and Lisa D. Brooks and Jean E. McEwen and Mark S. Guyer and Vivian Ota Wang and Jane L. Peterson and Michael Shi and Jack Spiegel and Lawrence M. Sung and Lynn F. Zacharia and Francis S. Collins and Karen Kennedy and Ruth Jamieson and John Stewart},
   doi = {10.1038/nature06258},
   issn = {00280836},
   issue = {7164},
   journal = {Nature},
   month = {10},
   pages = {851-861},
   pmid = {17943122},
   title = {A second generation human haplotype map of over 3.1 million SNPs},
   volume = {449},
   year = {2007},
}

@article{Liu2021,
   author = {Jared Liu and Hsin Wen Chang and Zhi Ming Huang and Mio Nakamura and Sahil Sekhon and Richard Ahn and Priscila Munoz-Sandoval and Shrishti Bhattarai and Kristen M. Beck and Isabelle M. Sanchez and Eric Yang and Mariela Pauli and Sarah T. Arron and Wai Ping Fung-Leung and Ernesto Munoz and Xuejun Liu and Tina Bhutani and Jeffrey North and Anne M. Fourie and Michael D. Rosenblum and Wilson Liao},
   doi = {10.1016/j.jaci.2020.11.028},
   issn = {10976825},
   issue = {6},
   journal = {Journal of Allergy and Clinical Immunology},
   month = {6},
   pages = {2370-2380},
   pmid = {33309739},
   publisher = {Mosby Inc.},
   title = {Single-cell RNA sequencing of psoriatic skin identifies pathogenic Tc17 cell subsets and reveals distinctions between CD8+ T cells in autoimmunity and cancer},
   volume = {147},
   year = {2021},
}

@article{Izar2020,
   author = {Benjamin Izar and Itay Tirosh and Elizabeth H. Stover and Isaac Wakiro and Michael S. Cuoco and Idan Alter and Christopher Rodman and Rachel Leeson and Mei Ju Su and Parin Shah and Marcin Iwanicki and Sarah R. Walker and Abhay Kanodia and Johannes C. Melms and Shaolin Mei and Jia Ren Lin and Caroline B.M. Porter and Michal Slyper and Julia Waldman and Livnat Jerby-Arnon and Orr Ashenberg and Titus J. Brinker and Caitlin Mills and Meri Rogava and Sébastien Vigneau and Peter K. Sorger and Levi A. Garraway and Panagiotis A. Konstantinopoulos and Joyce F. Liu and Ursula Matulonis and Bruce E. Johnson and Orit Rozenblatt-Rosen and Asaf Rotem and Aviv Regev},
   doi = {10.1038/s41591-020-0926-0},
   issn = {1546170X},
   issue = {8},
   journal = {Nature Medicine},
   month = {8},
   pages = {1271-1279},
   pmid = {32572264},
   publisher = {Nature Research},
   title = {A single-cell landscape of high-grade serous ovarian cancer},
   volume = {26},
   year = {2020},
}

@misc{AutoMiss,
   author = {Michael Chan},
   publisher = {GitHub},
   title = {auto-miss},
   url = {https://github.com/ClimerLab/auto-miss.},
   year = {2019},
}

@article{Bo2004,
   author = {T. H. Bo},
   doi = {10.1093/nar/gnh026},
   issn = {1362-4962},
   issue = {3},
   journal = {Nucleic Acids Research},
   month = {2},
   pages = {34e-34},
   title = {LSimpute: accurate estimation of missing values in microarray data with least squares methods},
   volume = {32},
   year = {2004},
}

@article{Oba2003,
   author = {Shigeyuki Oba and Masa-aki Sato and Ichiro Takemasa and Morito Monden and Ken-ichi Matsubara and Shin Ishii},
   doi = {10.1093/bioinformatics/btg287},
   issn = {1367-4811},
   issue = {16},
   journal = {Bioinformatics},
   month = {11},
   pages = {2088-2096},
   title = {A Bayesian missing value estimation method for gene expression profile data},
   volume = {19},
   year = {2003},
}

@article{Kim2005,
   author = {Hyunsoo Kim and Gene H. Golub and Haesun Park},
   doi = {10.1093/bioinformatics/bth499},
   issn = {1367-4811},
   issue = {2},
   journal = {Bioinformatics},
   month = {1},
   pages = {187-198},
   title = {Missing value estimation for DNA microarray gene expression data: local least squares imputation},
   volume = {21},
   year = {2005},
}

@article{Troyanskaya2001,
   author = {Olga Troyanskaya and Michael Cantor and Gavin Sherlock and Pat Brown and Trevor Hastie and Robert Tibshirani and David Botstein and Russ B. Altman},
   doi = {10.1093/bioinformatics/17.6.520},
   issn = {1367-4811},
   issue = {6},
   journal = {Bioinformatics},
   month = {6},
   pages = {520-525},
   title = {Missing value estimation methods for DNA microarrays},
   volume = {17},
   year = {2001},
}

@article{Musil2002,
   author = {Carol M. Musil and Camille B. Warner and Piyanee Klainin Yobas and Susan L. Jones},
   doi = {10.1177/019394502762477004},
   issn = {0193-9459},
   issue = {7},
   journal = {Western Journal of Nursing Research},
   month = {11},
   pages = {815-829},
   title = {A Comparison of Imputation Techniques for Handling Missing Data},
   volume = {24},
   year = {2002},
}

@article{Lin2020,
   author = {Wei-Chao Lin and Chih-Fong Tsai},
   doi = {10.1007/s10462-019-09709-4},
   issn = {0269-2821},
   issue = {2},
   journal = {Artificial Intelligence Review},
   month = {2},
   pages = {1487-1509},
   title = {Missing value imputation: a review and analysis of the literature (2006–2017)},
   volume = {53},
   year = {2020},
}

@book{Saaty1970,
   author = {Thomas L. Saaty},
   doi = {10.2307/2316512},
   editor = {Donald K. Prentiss},
   issn = {00029890},
   pages = {1-295},
   publisher = {McGraw-Hill, Inc},
   title = {Optimization in Integers and Related Extremal Problems.},
   year = {1970},
}
\end{document}